\documentclass[10pt,conference,letterpaper]{IEEEtran}
\IEEEoverridecommandlockouts 
\ifCLASSINFOpdf
  \usepackage[pdftex]{graphicx}
  \graphicspath{{figures/}}
\else
\fi

\usepackage{amsmath}
\usepackage{amssymb}
\usepackage{amsthm}
\usepackage{algorithm}
\usepackage{algorithmicx}
\usepackage{algpseudocode}
\usepackage{comment}
\usepackage{caption}
\usepackage{setspace}
\usepackage{color}
\newtheorem{theorem}{Theorem}

\newtheorem{definition}{Definition}

\newtheorem{lemma}{Lemma}
\usepackage[long]{optional}

\def\eg{\textit{e.g.}}
\def\ie{\textit{i.e.}}

\begin{document}
%
\title{Online Job Scheduling in Distributed Machine Learning Clusters}

\author{\IEEEauthorblockN{Yixin Bao\IEEEauthorrefmark{1},
Yanghua Peng\IEEEauthorrefmark{1},
Chuan Wu\IEEEauthorrefmark{1},
Zongpeng Li\IEEEauthorrefmark{2}\opt{short}{\thanks{This work was supported in part by grants from Hong Kong RGC under the contracts HKU 17204715, 17225516, C7036-15G (CRF), grants NSFC 61628209 and NSFC 61571335, HKU URC Matching Funding, and Hubei Science Foundation 2016CFA030, 2017AAA125.}}}
\IEEEauthorblockA{\IEEEauthorrefmark{1}Department of Computer Science, The University of Hong Kong, Email: \{yxbao,yhpeng,cwu\}@cs.hku.hk}
\IEEEauthorblockA{\IEEEauthorrefmark{1}Department of Computer Science, University of Calgary, Email: zongpeng@ucalgary.ca}}

\maketitle

\begin{abstract}
	Nowadays large-scale distributed machine learning systems have been deployed to support various analytics and intelligence services in IT firms. To train a large dataset and derive the prediction/inference model, e.g., a deep neural network, multiple workers are run in parallel to train partitions of the input dataset, and update shared model parameters. In a shared cluster handling multiple training jobs, a fundamental issue is how to efficiently schedule jobs and set the number of concurrent workers to run for each job, such that server resources are maximally utilized and model training can be completed in time. Targeting a distributed machine learning system using the parameter server framework, we design an online algorithm for scheduling the arriving jobs and deciding the adjusted numbers of concurrent workers and parameter servers for each job over its course, to maximize overall utility of all jobs, contingent on their completion times. Our online algorithm design utilizes a primal-dual framework coupled with efficient dual subroutines, achieving good long-term performance guarantees with polynomial time complexity. Practical effectiveness of the online algorithm is evaluated using trace-driven simulation and testbed experiments, which demonstrate its outperformance as compared to commonly adopted scheduling algorithms in today's cloud systems. 

\end{abstract}
\section{Introduction}

Most leading IT companies have deployed distributed machine learning (ML) systems, which train various machine learning models over large datasets for providing AI-driven services. For example, Google uses its scalable ML framework, TensorFlow, to power products such as Google Photos and Google Cloud Speech \cite{abadi2016tensorflow}. Microsoft employs its distributed cognitive toolkit, CNTK, for speech recognition and image related learning tasks \cite{Microsoft_ML}. Baidu developed a PArallel Distributed Deep LEarning (PaddlePaddle) system and extensively uses large-scale ML for advertising, group shopping, etc.~\cite{PaddlePaddle}. \opt{long}{Tencent has applied its large-scale ML system, Angel \cite{Tencent_Angel}, for social advertising, user portrait mining and other recommendation services. }In these scenarios, large ML clusters with hundreds or thousands of (GPU) servers are deployed, where many internal/external training jobs are run to derive various prediction/inference models, \eg, Deep Neural Networks (DNNs), Logistic Regression (LR), and Latent Dirichlet Allocation\opt{long}{ (LDA)}.

Training machine learning models is typically resource intensive and time consuming. For example, it takes $23.4$ hours to train a GoogLeNet model 
using the ImageNet dataset on a Titan supercomputer server with 32 NVIDIA K20 GPUs \cite{deng2009imagenet}\cite{iandola2016firecaffe}. 
A fundamental challenge faced by an ML cluster operator is how to efficiently schedule submitted training jobs to maximally exploit available server resources (especially the expensive GPU cards), and to complete training in an expedited fashion. In representative distributed ML systems \cite{abadi2016tensorflow}\cite{Microsoft_ML}\cite{PaddlePaddle}\cite{chen2016mxnet}, training is done in parallel by multiple concurrent {\em workers}. There are two parallelism models: {\em data parallelism}, where the input dataset is partitioned among the workers, and each worker has a local copy of the entire ML model, computes model parameter changes using allocated data chunks, and exchanges computation results with other workers to come up with the right global parameter updates \cite{li2014scaling}\cite{chen2016mxnet}; {\em model parallelism}, where the ML model is partitioned among workers and each worker updates part of the parameters using the entire dataset \cite{chilimbi2014project}. Data parallelism has been more widely adopted than model parallelism, given that most ML models can be entirely stored in the memory of modern GPUs, eliminating the need for \opt{short}{model partition}\opt{long}{partitioning a model}. For example, latest NVIDIA GPU models (TITAN X and Tesla) 
have a memory of \opt{long}{$12$GB, }$16$GB or $24$GB, sufficient for most state-of-the-art models (\eg, \cite{he2016deep}\cite{simonyan2015very}). 
We focus on data \opt{long}{parallel training jobs in this work.}\opt{short}{parallelism in this work.}

A typical approach to exchange parameter changes among workers is through a parameter server framework \cite{li2014scaling}\cite{chilimbi2014project}: There are one or multiple {\em parameter servers} (typically implemented as virtualized instances using virtual machines or containers), and model parameters are evenly divided and maintained by the parameter servers. In each training iteration, a worker sends its computed parameter changes to the parameter servers; the parameter servers update their maintained parameters respectively, and send updated parameters back to the worker. The number of concurrent workers, as well as the number of parameter servers to support parameter exchange
, decide the training speed and completion time of a job.

How are training jobs scheduled 
in the existing ML systems? Google uses Borg \cite{verma2015large} as the ML cluster scheduler. Microsoft, Tencent, and Baidu use customized versions of YARN-like schedulers \cite{vavilapalli2013apache} for managing ML jobs, based on our exchanges with their employees\opt{long}{ (there is little open discussion available)}. 
The default scheduling policies of these schedulers are typically FIFO (as in Spark \cite{zaharia2010spark}
), Dominant Resource Fairness Scheduling \cite{ghodsi2011dominant} (as in YARN \cite{vavilapalli2013apache} and Mesos \cite{hindman2011mesos}), or priority-based greedy approaches (as in Borg \cite{verma2015large}). To our knowledge, none of these systems allow a varying number of concurrent workers in a training job, which is specified by the job owner and remains fixed throughout the training course. Such static resource allocation to jobs may not fully utilize the (often expensive) ML cluster resources, preventing the best training speeds.

We propose an online job scheduling algorithm, tailored for operating a shared ML cluster running multiple training jobs. The algorithm, referred to as {\em OASiS}, computes the best job execution schedule upon the arrival of each job, based on projected resource availability in the future course and potential job utility to achieve (contingent on its completion time). Judging whether the potential job utility outweighs resource consumption, the algorithm decides admitting the job or not, and runs the job according to the best schedule if admitted. With the schedule, the numbers of workers and parameter servers and their deployment on servers are dynamically adjusted during the course of the job, for expedited training adapting to resource availability at different times. 
Over the long run, we seek overall \opt{short}{job }utility maximization\opt{long}{ of all training jobs}. 

Our online algorithm design utilizes an online primal-dual framework coupled with dual subroutines, to efficiently tackle the combinatorial online optimization problem. Based on the primal-dual framework, we maintain meticulously computed (dual) resource prices 
according to time-varying resource consumption levels (less resources when new jobs are admitted and more when jobs are completed), and decide job admission and resource allocation accordingly. 
Given the resource prices, the dual subroutines include efficient, optimal algorithms to compute the best schedule of worker and parameter server deployment for each job, exploiting a dynamic programming structure of the underlying multi-timeslot multi-dimensional resource packing problem.
 

We rigorously prove polynomial running time of our online algorithm, and its long-term performance guarantee in terms of a good competitive ratio in total job utility. We evaluate practical effectiveness of {\em OASiS} using trace-driven simulation and testbed experiments, by implementing it as a new scheduler module in Kubernetes \cite{Kubernetes_blog} for MXNet -- a popular distributed machine learning platform \cite{chen2016mxnet}. 
The results show that {\em OASiS} outperforms commonly adopted scheduling policies\opt{long}{ especially in systems with resource scarcity}. 

\section{Related Work}\label{relatedwork}
\subsection{Distributed Machine Learning Systems}

A number of distributed ML frameworks have been designed and deployed, \eg, TensorFlow \cite{abadi2016tensorflow}, CNTK \cite{Microsoft_ML}, PaddlePaddle \cite{PaddlePaddle}, 
MXNet \cite{chen2016mxnet}. 
The parameter server framework, mainly due to Li {\em et al.}~\cite{li2014scaling}, has been incorporated in some of them (\eg, \cite{chen2016mxnet}\cite{chilimbi2014project}). 
In these systems, a static set of workers are employed; new workers are deployed only upon failure of existing ones. Most adopt Borg or YARN-like schedulers for ML cluster management \cite{verma2015large}\cite{vavilapalli2013apache}.



Recently in the literature, Dorm \cite{sun2017towards} advocates partitioning an ML cluster, runs one ML application per partition, and dynamically resizes the partitions for resource efficiency and fairness, by solving a mixed integer linear program (MILP) using a standard solver. In comparison, we design an online algorithm to guide resource allocation over time with proven performance. 
Dolphin \cite{lee2016dolphin} solves a cost-minimizing problem to find an optimal number of nodes to use for an ML job, and reconfigures the system dynamically. It focuses on runtime optimization of one ML job, instead of optimal resource allocation among multiple concurrent jobs. Similarly, Yan {\em et al.}~\cite{yan2015performance} develop performance models to quantify the impact of model and data partitioning and system provisioning on training performance of a DNN, 
where online job scheduling and resource sharing are not considered.

\subsection{Job Scheduling and Resource Allocation in Cloud Systems}


\opt{long}{There have been many studies on admission control and job scheduling/resource allocation in general cloud systems. Rayon \cite{curino2014reservation} performs online admission control by accepting all jobs that can fit in the cluster agenda and rejecting ones that it can not satisfy, considering reservation of future resources. YARN \cite{vavilapalli2013apache} uses admission control to delay allocating fallow cluster resources to protect its own availability and schedules admitted jobs using a dominant resource fairness strategy. 
Apollo \cite{boutin2014apollo} utilizes various admission control policies and decides how and when to assign its resource quotas to submitted jobs in a virtual cluster, using estimation-based scheduling, {\em i.e.}, minimizing estimated task completion time by considering relevant factors historically. 
In comparison, we maintain virtual resource prices to decide job admission, which together with optimal resource scaling, achieves long-term overall job utility maximization.}

In the offline setting, Huang {\em et al.}~\cite{huang2015need} and Chen {\em et al.}~\cite{chen2017scheduling} study cloud job scheduling problems, targeting max-min fairness among jobs. 
For online scheduling, Azar {\em et al.}~\cite{azar2015truthful} propose an online preemptive job scheduling algorithm achieving a constant competitive ratio, for jobs running on a single machine with constant job utility. Lucier {\em et al.}~\cite{lucier2013efficient} propose an efficient heuristic for online job scheduling with preemption, aiming to maximize total value of all jobs. The resources allocated to each job are fixed over time and the job value 
is not influenced by completion time. Zhou {\em et al.}~\cite{zhou2017efficient} and Zhang {\em et al.}~\cite{xxzhangsigmetrics15} design mechanisms for online cloud resource allocation and pricing, where no adjustment of allocated resources in a job is considered.

Xiao {\em et al.}~\cite{xiao2014automatic} design a scheduler for automatic scaling of Internet applications in a cloud, targeting high demand satisfaction ratio and short request-response time. 
TetriSched \cite{tumanov2016tetrisched} enables resource scaling by periodically solving a schedule optimization problem among all pending jobs to compute their amounts of resources in need. 
These work do not provide theoretical guarantee for long-term performance. 


%
%
%
\section{Problem Model}\label{model}

\subsection{Distributed Machine Learning System}
\begin{figure}[!t]
\captionsetup{width=0.5\textwidth}
\centering
  \includegraphics[width=3.3in]{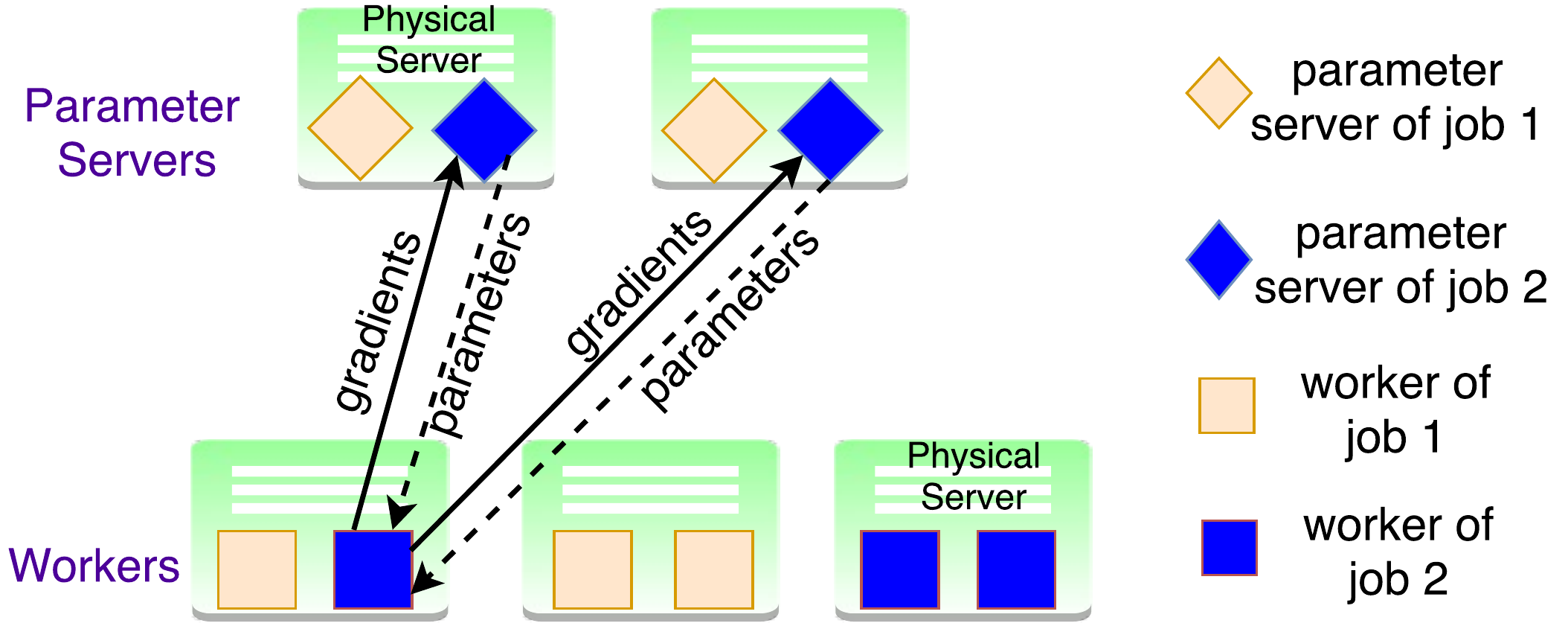}
  \caption{The distributed machine learning system}
  \label{paraserver}
\vspace{-6mm}
\end{figure}

Fig.~\ref{paraserver} illustrates an ML 
cluster, where a set of $I$ training jobs are submitted in an online fashion during timespan $1,2,\ldots, T$. The training jobs come with large input datasets, and derive potentially different ML models using {\em data parallel} training and the parameter server framework \cite{li2014scaling}. A job $i$ arrives at time $a_i\in[T]$,\footnote{We define $[X]=\{1,2,\ldots,X\}$ throughout the paper, where $X$ can be different quantities.} using a number of workers and parameter servers for model training. 

\captionsetup[figure]{labelfont=bf}

Workers and parameter servers are implemented on virtual machines (VMs) or containers in the physical servers.
The ML cluster hosts $H$ physical servers for worker deployment. Each machine $h\in [H]$ has a capacity $c_h^r$ of type-$r$ resource. There are $K$ other physical servers for running parameter servers, and each server $k\in [K]$ has a capacity $c_k^r$ of type-$r$ resource. Let $R$ be the total number of resource types, including GPU, CPU, memory, disk storage and bandwidth capacity of the server NIC. 
\opt{long}{We ignore disk IO constraint as SSDs are widely used in ML clusters and the data read/write delay is often negligible. }
We practically assume two types of physical machines for running workers and parameter servers separately, given that parameter servers are typically placed on machines with high bandwidth but without GPU, while workers run on GPU servers. Such a separation between workers and parameter servers has been witnessed in existing ML systems \cite{li2014scaling}\cite{chilimbi2014project}\opt{long}{\cite{ovtcharov2015accelerating}}. 

Workers and parameter servers are customized for each job, and not shared among different jobs. Each worker (parameter server) of job $i$ occupies a $w_i^r$ ($s_i^r$) amount of type-$r$ resource, $\forall r\in[R]$. 
An amount of bandwidth $b_i$ ($B_i$) is reserved for each worker (parameter server) of job $i$, \ie, $b_i=w_i^{\mbox{\small bandwidth}}$ ($B_i=s_i^{\mbox{\small bandwidth}}$). 
We do not distinguish upload and download bandwidth, but assume they are symmetric. Bandwidth reservation for a VM or container 
is common for accelerated computing in cloud platforms, to guarantee data transfer performance of each instance, \eg, the 
reserved bandwidth of EC2 GPU instance P2 on AWS is $10$Gbps or $20$Gbps \cite{amazongpuserver}.


\subsection{Asynchronous Training Workflow}

The input dataset to a training job is stored in a distributed storage system (\eg, HDFS \cite{hdfs}). The dataset is divided into equal-sized {\em data chunks} trained by different workers.\opt{long}{\footnote{We assume data chunks are assigned to workers based on a locality: the data chunks are stored in a HDFS-like distributed file system; each data chunk is assigned to workers based on the preference order of workers on the same server where there is a replica of the chunk, workers on the same rack with a replica, and other workers.}} 
Each data chunk is further divided into equal-sized {\em mini-batches}. 

Upon start, a worker fetches a data chunk.\opt{short}{\footnote{The ML framework, \eg, PaddlePaddle, assigns data chunks to workers.}} Then the worker processes the first mini-batch in the data chunk, \ie, computes what changes to be made to the parameters (to approach their optimal values) in the ML model, using data in the mini-batch. Parameter changes are typically expressed as {\em gradients} (directions of changes), 
and a distributed stochastic gradient descent method is typically used by workers to jointly improve the parameters \cite{li2014scaling}. For example, when training an LR model for ad click-through-rate prediction, parameters are the weights of features (\eg, text, image used in an ad) in the prediction model, and gradients are the changes of weights \cite{adclickprediction}. 

After processing a mini-batch, the worker sends gradients to the parameter servers for parameter updates. 
The parameter servers in a job are usually responsible for an evenly divided share of the parameters. In the above example, if there are two parameter servers, each will be responsible for half of the weights, 
and gradients computed by a worker are divided and sent to parameter servers maintaining respective weights. Upon receiving updated parameters from all parameter servers, the worker continues computing gradients using the next mini-batch, and so on. After an entire data chunk is processed, the worker continues training the next data chunk assigned to it.

\begin{figure}[!t]
\captionsetup{width=0.48\textwidth}
\centering
  \includegraphics[width=3.3in]{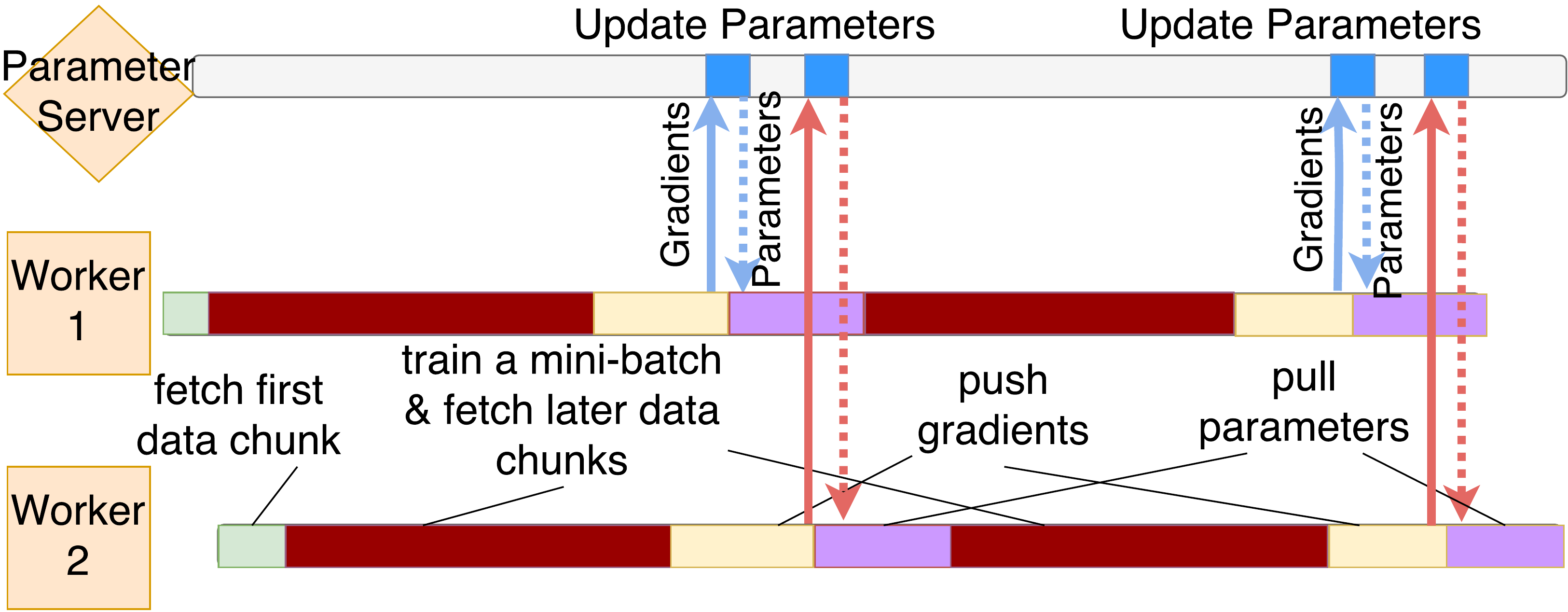}
  \caption{Workflow in a training job}
  \label{workflow}
\vspace{-8mm}
\end{figure}

Fig.~\ref{workflow} illustrates the {\em asynchronous training} workflow in our system, \ie, the training progress at different workers in a job is not synchronized and each parameter server updates its parameters each time upon receiving gradients from a worker. In the above example, a parameter server updates its weights using a formula like $new~weight=old~weight~-~step size~\times~gradient~computed~by~the~worker$, 
and then sends updated weights back to the worker. Another representative training mode in today's ML systems is synchronous training, where training progress at all workers is synchronized and each parameter server updates its parameters after it has collected gradients from all workers in each training iteration (\ie, training of one mini-batch). Asynchronous training achieves better bandwidth utilization, as gradients and updated parameters are sent from/to workers at different times, and hence potentially faster convergence. 
Further, 
model accuracy achieved with asynchronous training is not affected by changes of worker population through the course \cite{li2014scaling}\cite{chilimbi2014project} (as what we advocate), while it varies with synchronous training if different numbers of concurrent workers are used \cite{iandola2016firecaffe}\cite{yan2015performance}.

\captionsetup[figure]{labelfont=bf}

Let $N_i$ be the number of input data chunks in job $i$, each divided into $M_i$ mini-batches. Let $\tau_i$ denote the training time (gradient computation) for each mini-batch in job $i$, 
which is assumed to be equal for all mini-batches on all workers in the same job, given the same resource allocation per worker. 
Let $e_i$ be the size of gradients produced by each worker of job $i$ after processing a mini-batch, which is the same as the size of updated parameters that the worker will receive from all parameter servers, since the total numbers of gradients and parameters are always the same and both use the same float point representation \cite{iandola2016firecaffe}. 
The time for sending gradients to or receiving updated parameters from all parameter servers can be computed as $\frac{e_i}{b_i}$ (bandwidth at a parameter server is typically large enough to receive gradients /send parameters from/to multiple workers). 
When training ResNet-152 model 
on ImageNet dataset 
\cite{he2016deep}\cite{deng2009imagenet}, each data chunk is \opt{long}{typically }$128$MB in size, 
a mini-batch is about $6$MB in size, and training one mini-batch takes about one second, while training a data chunk takes less than one minute; the size of gradients/parameters exchanged between a worker and parameter servers is about $241$MB.

We ignore worker/parameter server setup time, since the image containing the training program can be pre-stored in a physical machine or fetched in a very short time (\eg, a container image of hundreds of MBs can be fetched within seconds in a 10Gbps network). 
We also ignore the time for a worker to fetch data chunks from distributed storage, since a worker only needs to explicitly retrieve the first chunk, 
and fetching time of later chunks can be hidden behind training through pipelining. Fetching one data chunk takes much shorter time than training, \eg, less than $1$s in a 10Gbps network for a $128$MB chunk. 
With asynchronous training, the computation time at a parameter server for updating parameters using gradients from only one worker is very small (around tens of milliseconds in ResNet-152
) and hence negligible too. 
 

In an ML job, input data chunks can be repeatedly trained for multiple rounds. An {\em epoch} \cite{chilimbi2014project} is the duration when all data chunks are trained once
. A training job $i$ stops after $E_i$ epochs in our system.

\subsection{Offline Optimization Problem}

Upon arrival of an ML job $i$ at $a_i$, the following decisions are made:\opt{long}{\footnote{We focus on internal ML jobs in a company such that the number of workers and parameter servers can be specified by our algorithm.}} (i) Whether the job should be admitted, denoted by a binary variable $x_i$: $x_i=1$ if job $i$ is admitted, and $x_i=0$, otherwise. Admission control is common in cloud management systems \cite{verma2015large}\cite{vavilapalli2013apache}, 
and jobs that are not admitted can be queued or resubmitted at a later time beyond $T$. 
(ii) The number of workers of job $i$ to deploy on physical server $h\in[H]$ in each time slot at and after $a_i$, indicated by integer variable $y_{ih}(t)$. 
(iii) The number of parameter servers of job $i$ to deploy on physical server $k\in[K]$ in each time slot at and after $a_i$, denoted by integer variable $z_{ik}(t)$. 

Given that it is not practical to adjust worker and parameter server deployment frequently, the length of each time slot 
is potentially much larger than the duration of an epoch. For example, one time slot can be 1 hour or longer. 

Let $\hat{t}_i$ be the completion time slot of job $i$. Each job $i$ has a non-negative utility $f_i(\hat{t}_i-a_i)$, non-increasing with $\hat{t}_i-a_i$, specifying the job's value in terms of different completion times \cite{huang2015need}\opt{long}{\cite{chen2016optimizing}}. 
The offline optimization problem to maximize overall utility is formulated as follows. Important notation is summarized in Table \ref{table1}.
 
 \vspace{-1mm}
 {\small
\begin{equation}
	\max \sum_{i \in [I]}x_if_i(\hat{t}_i-a_i )\label{obj}
\end{equation}

\vspace{-3mm}
subject to:	
\vspace{-5mm}

\begin{align}
& \sum_{t \in [T]} \sum_{h \in [H]} y_{ih}(t) \geq E_i N_i M_i(\tau_i + 2e_i/b_i)x_i, \forall i \in [I]\label{constraint1}\\ 
& \sum_{h \in [H]} y_{ih}(t)\leq N_i x_i, \forall i \in [I], t \in [T]:t \geq a_i \label{constraint2}\\ 
& \sum_{i \in [I]}  w_i^r y_{ih}(t) \leq c_h^r, \forall t\in[T],r \in [R],h \in [H]\label{constraint4}\\
& \sum_{i \in [I]}  s_i^r z_{ik}(t) \leq c_k^r, \forall t\in[T],r \in [R],k \in [K]\label{constraint5}\\
& \sum_{h \in [H]} y_{ih}(t) b_i \leq \sum_{k\in[K]} z_{ik}(t) B_i,\forall i \in [I],t\in[T]\label{constraint8}\\
& \sum_{k\in[K]} z_{ik}(t) \leq \sum_{h \in [H]} y_{ih}(t),\forall i \in [I],t\in[T]\label{constraint9}\\
& \hat{t}_i =\arg\max_{t\in[T]} \{\sum_{h\in[H]}y_{ih}(t)>0\},\forall i\in[I]\label{constraint7}\\
& y_{ih}(t)=0,\forall i\in[I],h\in[H],t<a_i\label{constraint10}\\
& z_{ik}(t)=0,\forall i\in[I], k\in[K],t<a_i\label{constraint11}\\
& x_i \in \{0,1\}, \forall i\in[I] \label{constraint12}\\
& y_{ih}(t)\in\{0,1, \ldots\}, \forall i\in[I],t\in[T],h\in[H]\label{constraint13}\\
& z_{ik}(t)\in\{0,1, \ldots\}, \forall i\in[I],t\in[T],k\in[K] \label{constraint14}
\end{align}
}
\vspace{-2mm}

Constraint (\ref{constraint1}) ensures that for each admitted job $i$, a sufficient number of workers are deployed to accomplish training of its dataset for $E_i$ epochs. Here, $\tau_i + 2e_i/b_i$ is the time for training a mini-batch, sending gradients to parameter servers, and receiving updated parameters from parameter servers. 
$E_i N_i M_i$ is the total count of mini-batches trained in the job. $\sum_{t \in [T]} \sum_{h \in [H]} y_{ih}(t)$ indicates the total amount of work time that all deployed workers in job $i$ provide. 
(\ref{constraint2}) specifies the concurrent number of workers of job $i$ should be no more than the number of data chunks $N_i$, to ensure that one data chunk is processed by at most one worker in each time slot (such that data chunks are trained evenly over time). 
(\ref{constraint4}) and (\ref{constraint5}) are resource capacity constraints on physical machines for worker and parameter server deployment, respectively. 
(\ref{constraint8}) guarantees that the total bandwidth of parameter servers is no smaller than total bandwidth of all workers in each job, \ie, parameter servers will not be bottlenecks during gradient/parameter exchange. 
(\ref{constraint9}) upper bounds the number of parameter servers by the number of workers at any time in each job, which is common in practical ML systems \cite{li2014scaling}\cite{chilimbi2014project}. 
(\ref{constraint7}) gives the completion time slot of job $i$. 
(\ref{constraint10}) and (\ref{constraint11}) set worker and parameter server numbers to 0 before a job's arrival. 
 





The optimization problem involves integer variables and non-conventional constraints in (\ref{constraint7}). We design an efficient online algorithm to solve it in an online fashion, without assuming knowledge of any future job arrivals. 

\begin{table}[!t] 
\renewcommand{\arraystretch}{1.3}
\caption{Notation}\label{table1}
{\small
\begin{center}
\begin{tabular}{|l|l|l|l|}
\hline
{\small $I$} & {\small $\#$ of jobs} & {\small $T$} & {\small system timespan}\\ \hline
$\hat{t}_i$ & completion time of job $i$ & $a_{i}$ & arrival time of job $i$\\ \hline
$R$ & $\#$ of resource types & $N_i$ & $\#$ of data chunks in $i$\\ \hline
$x_i$ & accept job $i$ or not & $f_i(\cdot)$ & job $i$'s utility\\ \hline
\end{tabular}
\begin{tabular}{| p{0.85cm} | p{7cm} |}
$E_i$ & $\#$ of training epochs for job $i$\\ \hline
$M_i$ & $\#$ of mini-batches in a data chunk of job $i$\\ \hline
{\small $H$($K$)} & $\#$ of servers to deploy workers (parameter servers)\\ \hline
$c_h^r (c_k^r)$ & capacity of type-$r$ resource on server $h$ ($k$) to deploy workers (parameter servers)\\ \hline
$w_i^r (s_i^r)$ & type-$r$ resource of a worker (parameter server) in $i$\\ \hline
$y_{ih}(t)$ & $\#$ of workers of job $i$ deployed on server $h$ in $t$\\ \hline
$z_{ik}(t)$ & $\#$ of parameter servers of $i$ deployed on server $k$ in $t$\\ \hline
$b_i (B_i)$ & bandwidth of a worker (parameter server) of job $i$\\ \hline
$\tau_i$ & time to train a mini-batch in job $i$\\ \hline
$e_i$ & size of gradients/parameters exchanged between a worker and parameter servers in job $i$\\ \hline
\opt{long}{
$x_{il}$ & select schedule $l$ for job $i$ or not\\ \hline
$t_{il}$ & the completion time slot of job $i$ with schedule $l$ \\ \hline
$y_{ih}^{l}(t)$ & $\#$ of workers on server $h$ in $t$ in schedule $l$ of job $i$\\ \hline
$z_{ik}^{l}(t)$ & $\#$ of parameter servers on server $k$ in $t$ in schedule $l$ of job $i$\\ \hline
$\mathcal{L}_i$ & the set of feasible schedules of job $i$\\ \hline
}
\end{tabular}
\end{center}}
\vspace{-8mm}
\end{table}

\vspace{-1mm}
\section{Online Algorithm}\label{algorithm}

\subsection{Problem Reformulation}

To circumvent the non-conventional constraint (\ref{constraint7}), we reformulate problem (\ref{obj}) into the following integer linear program (ILP). 
Here $\mathcal{L}_i$ is the set of feasible schedules for jobs $i$, each corresponding to the set of decisions $(y_{ih}(t),z_{ik}(t),\forall h\in[H], k\in[K],t\in[T])$ satisfying constraints (\ref{constraint1})(\ref{constraint2})(\ref{constraint8})(\ref{constraint9})(\ref{constraint10})-(\ref{constraint14}). There is potentially an exponential number of feasible schedules for each job, due to combinatorial nature of those constraints. Decision variables in the ILP are binary variables $x_{il}$, indicating whether job $i$ is admitted and scheduled according to schedule $l\in \mathcal{L}_i$ or not, $\forall i\in[I],l\in \mathcal{L}_i$. Job $i$'s completion time according to schedule $l$ is $t_{il}$. $y_{ih}^{l}(t)$ ($z_{ik}^{l}(t)$) is the given number of workers (parameter servers) on server $h$ ($k$) in $t$ in job $i$'s schedule $l$ (not decision variables in (\ref{outobj})).

{\small
\begin{equation}
	\max_{\mathbf{x}} \sum_{i \in [I]}\sum_{l\in \mathcal{L}_i}x_{il}f_{i}(t_{il}-a_i)\label{outobj}
\end{equation}

\vspace{-3mm}
s.t. 
\vspace{-4mm}

\begin{align}
& \sum_{i\in[I]}\sum_{l: t\in l, h\in (t,l)} w_i^r y_{ih}^l(t) x_{il}\leq c_h^r, \forall t\in[T],r\in[R],h\in[H]\label{outconstraint1}\\
& \sum_{i\in[I]}\sum_{l: t\in l, k\in (t,l)} s_i^r z_{ik}^l(t) x_{il}\leq c_k^r, \forall t\in[T],r\in[R],k\in[K]\label{outconstraint2}\\
& \sum_{l\in \mathcal{L}_i} x_{il}\leq 1,\forall i \in [I]\label{outconstraint3}\\ 
& x_{il} \in \{0,1\},\forall i\in[I],l\in \mathcal{L}_i \label{outconstraint4}
\end{align}
}
\vspace{-4mm}

\noindent We use $t\in l, h\in (t,l), k\in (t,l)$ to indicate that schedule $l$ 
uses server $h$ to deploy worker(s) and server $k$ to deploy parameter server(s) for job $i$ in $t$. 
(\ref{outobj}), (\ref{outconstraint1}) and (\ref{outconstraint2}) are equivalent to (\ref{obj}), (\ref{constraint4}) and (\ref{constraint5}), respectively. (\ref{outconstraint3}) and (\ref{outconstraint4}) correspond to (\ref{constraint1})(\ref{constraint2})(\ref{constraint8})-(\ref{constraint14}). Problems (\ref{obj}) and (\ref{outobj}) are equivalent since a feasible solution to (\ref{obj}) has a corresponding feasible solution to (\ref{outobj}), and vice versa, with the same objective values. 
Though the number of variables in (\ref{outobj}), $x_{il}$'s, is potentially exponential, 
we will design an efficient online algorithm to solve (\ref{outobj}) in polynomial time, exploiting the primal-dual framework \cite{buchbinder2009design}. 
We formulate the dual of (\ref{outobj}) by relaxing integrality constraints (\ref{outconstraint4}) 
and associating dual variables $p_h^r(t)$, $q_k^r(t)$ and $\mu_i$ with (\ref{outconstraint1}), (\ref{outconstraint2}) and (\ref{outconstraint3}), respectively. 

\vspace{-2mm}
{\small
\begin{equation}
	\min \sum_{i\in[I]} \mu_i + \sum_{t\in[T]}\sum_{h\in[H]}\sum_{r\in[R]} p_h^r(t)c_h^r + \sum_{t\in[T]}\sum_{k\in[K]}\sum_{r\in[R]} q_k^r(t)c_k^r\label{dualobj}
\end{equation}
 
\vspace{-5mm}
 
\begin{align}
s.t.\ & \mu_i \geq f_i(t_{il}-a_i) - \sum_{t \in l} \sum_{h \in (t,l)} \sum_{r\in[R]} p_h^r(t) w_i^r y_{ih}^{l}(t)\nonumber\\
&~~~~~ - \sum_{t \in l} \sum_{k \in (t,l)} \sum_{r\in[R]} q_k^r(t) s_i^r z_{ik}^{l}(t),\forall i\in[I],l \in \mathcal{L}_i\label{outdualconstraint1}\\
& p_h^r(t)\geq 0,q_k^r(t)\geq 0, \forall t\in[T],h\in[H],k\in[K],r\in[R]\nonumber\\
& \mu_i \geq 0,\forall i \in[I]\nonumber
\end{align}
}
\vspace{-4mm}

\noindent The dual variable $p_h^r(t)$ ($q_k^r(t)$), associated with the primal capacity constraint on server $h$ ($k$), can be interpreted as the unit cost for type-$r$ resource on the server in $t$. Then {\small$\sum_{t \in l} \sum_{h \in (t,l)} \sum_{r\in[R]} p_h^r(t) w_i^r y_{ih}^{l}(t)$} ({\small$\sum_{t \in l} \sum_{k \in (t,l)} \sum_{r\in[R]} q_k^r(t) s_i^r z_{ik}^{l}(t)$}) is the total resource cost of all workers (parameter servers) of job $i$ with schedule $l$. 
The RHS of (\ref{outdualconstraint1}) is job utility minus overall resource cost for job $i$ with schedule $l$. The following should hold to minimize the dual objective: 
$\mu_i =\max\{0,  \max_{l\in\mathcal{L}_i} \mbox{RHS of (\ref{outdualconstraint1})}\}$. 
\noindent Hence, $\mu_i$ can be nicely interpreted as the payoff of admitting job $i$ according to the best schedule $l^*$: 



\vspace{-5mm}
{\small
\begin{align}
	l^*=\arg\max_{l\in\mathcal{L}_i} \mbox{~~RHS of (\ref{outdualconstraint1})}\label{opt_schedule}
\end{align}
}
\vspace{-7mm}

\subsection{Online Algorithm}
These observations inspire the design of an online algorithm: Upon arrival of job $i$, we compute the best schedule $l^*$ of job $i$ (assuming job admitted). Then we check if the RHS of (\ref{outdualconstraint1}) achieved by $l^*$ is positive. If so ($\mu_i>0$, positive payoff), we accept job $i$ and run it according to $l^*$ ($x_{il^*}=1$); otherwise (zero payoff), job $i$ is rejected ($x_{il}=0,\forall l\in\mathcal{L}_i$). 
The rationale is that, as resources are limited, we wish to accept jobs with larger utility and lower resource consumption, to maximize (\ref{outobj}). 
A positive payoff indicates that the job utility is high enough to justify resource consumption, and we schedule the job in a way that maximizes its payoff.

To implement this idea, we need to resolve the following: (i) Solve 
(\ref{opt_schedule}) to find the best schedule $l^*$ for job $i$. Simply enumerating all feasible schedules is not practical, given the exponential size of set $\mathcal{L}_i$. We will design an efficient subroutine to produce $l^*$ in polynomial time in Sec.~\ref{subroutine}. (ii) Compute dual resource prices $p_h^r(t)$'s and $q_k^r(t)$'s, 
to ensure a positive payoff for job schedules achieving high utilities (if there are enough resources to accommodate them), and non-positive payoff for job schedules resulting in low utilities or without available resources.


\setlength{\textfloatsep}{3pt}
\begin{algorithm}[!t]
\caption{{\em OASiS}: Online Algorithm for Scheduling ML Jobs}
\label{algo1}
\algdef{SE}[EVENT]{Event}{EndEvent}[1]{\textbf{Upon}\ #1\ \algorithmicdo}{\algorithmicend\ \textbf{upon}}
{\small
\begin{algorithmic}[1]
   	\Require $T, c_h^r, c_k^r, \forall h \in [H], k \in [K], r\in[R]$
   	\Ensure $x_i, y_{ih}(t), z_{ik}(t),\forall i\in[I],t\in[T],h\in[H], k\in[K]$
	\State Initialize $y_{ih}(t)=0, z_{ik}(t)=0, g_h^r(t)=0,v_k^r(t)=0, p_h^r(t)=P_h^r(0), q_k^r(t)=Q_k^r(0),\forall i\in[I], t\in[T], h\in[H], k\in[K], r\in[R]$ \label{initialization}
	\Event{arrival of job $i$}
		\State Compute the best schedule $l^*$ and payoff $\mu_{i}$ for job $i$ using Alg.~\ref{algo2}\label{linecalloracle}
		\If {$\mu_{i}>0$}\label{linepospayoff}
			\State Set $x_i=1$\label{lineacjob}
			\State Set $y_{ih}(t),z_{ik}(t)$ according to schedule $l^*$, $\forall t\in l^*, h\in(t,l^*), k\in (t,l^*)$\label{lineacpolicy}
			\State Update $g_h^r(t)=g_h^r(t)+ w_i^r y_{ih}(t), \forall t \in l^*, h \in (t,l^*), r \in [R]$\label{lineupdatez}
			\State Update $p_h^r(t)=P_h^r(g_h^r(t)), \forall t\in l^*, h\in (t,l^*), r\in[R]$\label{lineupdatep}
			\State Update $v_k^r(t)=v_k^r(t)+ s_i^r z_{ik}(t), \forall t \in l^*, k \in (t,l^*), r \in [R]$\label{lineupdateh}
			\State Update $q_k^r(t)=Q_k^r(v_k^r(t)), \forall t\in l^*, k\in (t,l^*), r\in[R]$\label{lineupdateq}
			\State Accept and launch job $i$ according to schedule $l^*$
		\Else
			\State Set $x_i=0$ and reject job $i$\label{linerejjob}
		\EndIf
  	\EndEvent
\end{algorithmic}
}
\end{algorithm}

The sketch of our online algorithm, {\em OASiS}, is in Alg.~\ref{algo1}. In line \ref{linecalloracle}, Alg.~\ref{algo2} is the subroutine to compute $l^*$. In line \ref{lineupdatez} (\ref{lineupdateh}), $g_h^r(t)$ ($v_k^r(t)$) records the amount of allocated type-$r$ resource on server $h$ ($k$) for (future) time slot $t$. 
In lines \ref{lineupdatep} and \ref{lineupdateq}, we update dual resource prices using carefully designed price functions $P_h^r(\cdot)$ and $Q_k^r(\cdot)$, respectively: 

\vspace{-5mm}
{\small
\begin{align}
P_h^r(g_h^r(t))=L_1\Big(\frac{U_1^r}{L_1}\Big)^{\frac{g_h^r(t)}{c_h^r}},
~~~ Q_k^r(v_k^r(t))=L_2\Big(\frac{U_2^r}{L_2}\Big)^{\frac{v_k^r(t)}{c_k^r}}\label{marginalcosts}
\end{align}
}

\vspace{-8mm}

{\small
\begin{align}
\mbox{where} ~~& U_1^r=\max_{i\in[I]} \frac{f_i(\lceil {E_iM_i (\tau_i + 2e_i/b_i)} \rceil-a_i)}{w_i^r},\forall r\in[R]\label{U_1}
\end{align}

\begin{align}
& U_2^r=\max_{i\in[I]} \frac{f_i(\lceil {E_iM_i (\tau_i + 2e_i/b_i)} \rceil-a_i)}{s_i^r},\forall r\in[R]\label{U_2}\\
& L_1=\frac{1}{4\eta_1}\min_{i\in[I]} \frac{f_i( T-a_i)}{\sum_{r\in[R]} \lceil E_iN_iM_i (\tau_i + 2e_i/b_i) \rceil w_i^r}\label{L_1}\\
& L_2=\frac{1}{4\eta_2}\min_{i\in[I]} \frac{f_i(\lceil T-a_i)}{\sum_{r\in[R]} \lceil E_iN_iM_i (\tau_i + 2e_i/b_i) \rceil s_i^r}\label{L_2}
\end{align}
}
\vspace{-3mm}

\noindent $U_1^r$ ($U_2^r$) is the maximum per-unit-resource job utility for type-$r$ resource on physical servers to deploy workers (parameter servers), among all jobs. 
Here, $f_i(\lceil {E_iM_i (\tau_i + 2e_i/b_i)} \rceil-a_i)$ is the largest utility that job $i$ can achieve, by using the maximum number of workers ({\em i.e.}, $N_i$) at all times in $E_i$ training epochs to achieve the shortest job completion time $\lceil\frac{E_iN_iM_i (\tau_i + 2e_i/b_i)}{N_i} \rceil=\lceil {E_iM_i (\tau_i + 2e_i/b_i)} \rceil$. $L_1$ ($L_2$) represents the minimum unit-time-unit-resource job utility on physical servers to deploy workers (parameter servers), among all jobs. 
Here, $f_i(T-a_i)$ is the smallest utility that job $i$ may achieve, when it ends at $T$. $\eta_1$ and $\eta_2$ are scaling factors satisfying $\frac{1}{\eta_1} \leq \frac{\lceil E_iN_iM_i (\tau_i + 2e_i/b_i) \rceil \sum_{r\in[R]} w_i^r}{T\sum_{h\in[H]}\sum_{r\in[R]}c_h^r},\forall i\in[I]$, and $\frac{1}{\eta_2} \leq \frac{\lceil E_iN_iM_i (\tau_i + 2e_i/b_i) \rceil \sum_{r\in[R]} s_i^r}{T\sum_{k\in[K]}\sum_{r\in[R]}c_k^r},\forall i\in[I]$, to ensure the initial value of dual objective is bounded. 


The rationales behind our price functions are as follows. (i) The prices should be low enough at the beginning to accept many incoming jobs. When 
$g_h^r(t)=0$, $v_k^r(t)=0$, we have $p_h^r(t)=L_1,q_k^r(t)=L_2, \forall h\in[H], k\in[K],r\in[R]$, and then any job can be admitted at this point since $L_1$ and $L_2$ represent the lowest unit job utility (a formal proof is given in \opt{short}{our technical report \cite{techreport}}\opt{long}{Appendix \ref{proofmaginalcosts}}). (ii) The prices increase exponentially when the allocated amounts of resources increase, to filter out jobs with low utilities which arrive early, and to reserve resources for jobs with higher utilities that may arrive later. 
(iii) The respective price should be high enough when a resource on a server is exhausted, such that no more jobs requiring this resource are admitted. 
When $g_h^r(t)=c_h^r$ or $v_k^r(t)=c_k^r$, we have $p_h^r(t)=U_1^r$ or $q_k^r(t)=U_2^r$, and no more jobs requiring these resources would be admitted since $U_1^r$ and $U_2^r$ are the largest unit job utilities (proof in \opt{short}{\cite{techreport}}\opt{long}{Appendix \ref{proofmaginalcosts}}). The price functions are important to guarantee a good competitive ratio for our online algorithm. 

$U_1^r$, $U_2^r$, $L_1$ and $L_2$ are required to compute price functions in Alg.~\ref{algo1}, whose exact values are not known before all jobs have arrived. Instead, we adopt their estimated values (based on past experience) in our online algorithm, and will evaluate impact of inaccurate estimates in Sec.~\ref{evaluation}. 

\subsection{Subroutine for Finding Best Job Schedule}
\label{subroutine}

The optimization problem in (\ref{opt_schedule}) to compute the best schedule $l^*$ for job $i$ is equivalent to the following:


\vspace{-3mm}
{\small
\begin{align}
	\max_{\hat{t}_i,\mathbf{y},\mathbf{z}} f_i(\hat{t}_i-a_i) - \sum_{t\in[T]}\sum_{h\in[H]}\sum_{r\in[R]} p_h^r(t)w_i^r y_{ih}(t)\nonumber\\
	-  \sum_{t\in[T]}\sum_{k\in[K]}\sum_{r\in[R]} q_k^r(t)s_i^r z_{ik}(t)\label{dualoracle}
\end{align}

\vspace*{-4mm}
\begin{align}
s.t.\ & g_h^r(t) + w_i^r y_{ih}(t)\leq c_h^r,\forall t\in[T],r\in[R],h\in[H]\nonumber\\
& v_k^r(t) + s_i^r z_{ik}(t)\leq c_k^r,\forall t\in[T],r\in[R],k\in[K]\nonumber\\
& \mbox{Constraints (\ref{constraint1})(\ref{constraint2})(\ref{constraint8})-(\ref{constraint11})(\ref{constraint13})(\ref{constraint14}), where }x_i=1\nonumber
\end{align}
}
\vspace*{-5mm}

We next show that (\ref{dualoracle}) can be efficiently and optimally solved using dynamic programming and a greedy algorithm. When we fix $\hat{t}_i$, (\ref{dualoracle}) is simplified to the following ILP, where $\mathcal{T}_i=\hat{t}_i,\mathcal{D}_i=E_i N_i$:

\vspace{-5mm}
{\small
\begin{align}
 \min_{\mathbf{y},\mathbf{z}}~~~\mathbf{cost(\mathcal{T}_i, \mathcal{D}_i)}=&\sum_{t\in[a_i,\mathcal{T}_i]}\sum_{h\in[H]}\sum_{r\in[R]} p_h^r(t)w_i^r y_{ih}(t)  \nonumber\\
&+\sum_{t\in[a_i,\mathcal{T}_i]}\sum_{k\in[K]}\sum_{r\in[R]} q_k^r(t)s_i^r z_{ik}(t)\label{oracleilp}
\end{align}

\vspace{-4mm}
\begin{align}
s.t.\ &\sum_{t \in [a_i,\mathcal{T}_i]} \sum_{h \in [H]} y_{ih}(t) \geq  \mathcal{D}_i M_i(\tau_i + 2e_i/b_i) \label{oracleconstraint1}\\
& y_{ih}(t)\leq \min_{r\in[R]} \lfloor \frac{c_h^r - g_h^r(t)}{w_i^r}\rfloor,\forall h\in[H],t\in[a_i,\mathcal{T}_i]\label{oracleconstraint4}\\
& z_{ik}(t)\leq \min_{r\in[R]} \lfloor \frac{c_k^r - v_k^r(t)}{s_i^r}\rfloor,\forall k\in[K],t\in[a_i,\mathcal{T}_i]\label{oracleconstraint5}\\
& \mbox{(\ref{constraint2})(\ref{constraint8})(\ref{constraint9})(\ref{constraint13})(\ref{constraint14}), where } t\in[a_i,\mathcal{T}_i]\nonumber
\end{align}
}
\vspace{-5mm}

In problem (\ref{oracleilp}), deployment decisions in different time slots are coupled only in constraint (\ref{oracleconstraint1}), which requires sufficient workers and parameter servers to be deployed such that all $N_i$ data chunks are trained for $E_i$ epochs during $[a_i, \mathcal{T}_i]$. We refer to $\mathcal{D}_i$ in the RHS of (\ref{oracleconstraint1}) as {\em training workload}, indicating the total count of data chunks trained (a data chunk is counted $E_i$ times if trained for $E_i$ times). 
Since the time for training a data chunk is much smaller than the duration of a time slot, we may safely assume a worker trains an integer number of data chunks in each time slot. 
The training workload is distributed over different time slots in $[a_i, \mathcal{T}_i]$. If we know how much training workload (denoted by $\mathcal{D}_i(t)$) is to be fulfilled in a time slot $t$, we are left with a further simplified problem:

\vspace{-5mm}
{\small
\begin{align}
\min \mathbf{cost\_t}(t,\mathcal{D}_i(t))=\sum_{h\in[H]}\sum_{r\in[R]} p_h^r(t)w_i^r y_{ih}(t) \nonumber\\
+ \sum_{k\in[K]}\sum_{r\in[R]} q_k^r(t)s_i^rz_{ik}(t)\label{greedyilp}
\end{align}

\vspace{-4mm}
\begin{align}
s.t.\ &\sum_{h\in[H]}y_{ih}(t) \ge \mathcal{D}_i(t) M_i(\tau_i + 2e_i/b_i) \nonumber\\
& \mbox{(\ref{oracleconstraint4})(\ref{oracleconstraint5})(\ref{constraint2})(\ref{constraint8})(\ref{constraint9})(\ref{constraint13})(\ref{constraint14}), for the specific } t\nonumber
\end{align}
}
\vspace{-5mm}

Though (\ref{greedyilp}) is an ILP, it can be optimally solved using a greedy algorithm (to be discussed in Alg.~\ref{algo2} and analyzed in Theorem \ref{thoptdual}). Therefore, we come up with the following algorithm to find the best schedule for job $i$: enumerate end times $\hat{t}_i$ from $a_i$ to $T$; given $\hat{t}_i$, design a dynamic programming approach to compute how to best distribute the training workload over time slots in $[a_i, \hat{t}_i]$; then use the greedy algorithm to decide deployment of workers and parameter servers in each time slot. Our algorithm is given in Alg.~\ref{algo2}. 

\begin{algorithm}[!t]
\caption{Subroutine for Deriving Best Schedule of Job $i$}
\label{algo2}
{\small
\begin{algorithmic}[1]
   	\Require $T$, $p_h^r(t), g_h^r(t), q_k^r(t), v_k^r(t), c_h^r, c_k^r, \forall h\in[H], k \in [K], r\in[R], t \in[T]$
   	\Ensure best schedule $l^*$ and payoff $\mu_{i}$ for job $i$
	\State Initialize $\mu_{i}=0$, $l^*=\emptyset$, $y_{ih}(t)=0,z_{ik}(t)=0, \forall t\in[T], h\in[H], k\in[K]$
	\For{$\hat{t}_i=a_i$ to $T$}\label{lineenumtime}
		\State $(cost,l) =DP\_COST(\hat{t}_i, E_iN_i)$\label{lineilp}
		\State $\mu_{il}=f_i(\hat{t}_i-a_i) - cost$\label{mu_il}
		\If{$\mu_{il}>\mu_{i}$}\label{linesavebegin}
			\State $l^*\Leftarrow l$, $\mu_{i}=\mu_{il}$ 
		\EndIf\label{linesaveend}
	\EndFor
	\State return $l^*$, $\mu_{i}$\label{linereorgan}
	
	\vspace{2mm}	
	\Function{$\mathbf{DP\_COST}$}{$\mathcal{T}_i, \mathcal{D}_i$}\label{DP_cost_begin}
		\State $min\_cost=\infty$, $l=\emptyset$
		\For{$d=0$ to $\mathcal{D}_i$}\label{calcostbegin}
			\State $(cost\_t, \mathbf{y}(\mathcal{T}_i),\mathbf{z}(\mathcal{T}_i))=COST\_t(\mathcal{T}_i, d)$\label{costt}
			\State $(cost, l')=DP\_COST(\mathcal{T}_i-1, \mathcal{D}_i-d)$\label{dpcall}
			\If{$min\_cost>cost\_t+cost$}\label{compcostbegin}
				\State $min\_cost=cost\_t+cost$, $l\Leftarrow l'\cup \{\mathbf{y}(\mathcal{T}_i),\mathbf{z}(\mathcal{T}_i)\}$ 
			\EndIf\label{compcostend}
		\EndFor\label{calcostend}
		\State Return $min\_cost$, $l$
	\EndFunction\label{DP_cost_end}
	
	\vspace{2mm}
	\Function{$\mathbf{COST\_t}$}{$t, d$}\label{cost_t_begin}
		\State Initialize $y_{ih}(t)=0,z_{ik}(t)=0, \forall h\in[H], k\in[K]$
		\State Sort servers in $[H]$ according to $\sum_{r\in[R]}p_h^r(t)w_i^r$ in non-decreasing order into $h_1,h_2,\dots,h_H$\label{sortworker}
		\State $D=\lceil d M_i(\tau_i + 2e_i/b_i)\rceil$;\label{workload}
		\For{$j=1,\dots,H$}/*deploy workers*/
			\State $y_{ih_j}(t)=\min\Big\{\min_{r\in[R]} \lfloor \frac{c_h^r - g_h^r(t)}{w_i^r} \rfloor,$ 
			\State ~~~~~~~~~~~~~~~~~~~~$N_i- \sum_{j'=1}^{j-1} y_{ih_{j'}}(t), D\Big\}$\label{linesety}
			\State $D=D-y_{ih_j}(t)$
		\EndFor\label{workerdeployend}
	
		\If{$D>0$}/*not all workload can be handled*/\label{checkworkerstart}
			\State Return $cost\_t=+\infty, \mathbf{y}, \mathbf{z}$
		\EndIf\label{checkworkerend}
	
		\State Sort servers in $[K]$ according to $\sum_{r\in[R]}q_k^r(t)s_i^r$ in non-decreasing order into $k_1,k_2,\dots,k_K$\label{sortparaserver}
		\For{$j=1,\dots,K$}/*deploy parameter servers*/\label{paradeploystart}
			\State $z_{ik_j}(t)=\min\Big\{\min_{r\in[R]} \lfloor \frac{c_k^r - v_k^r(t)}{s_i^r} \rfloor,$
			\State ~~~~~~~~~~~~~~~~~~~~$\lceil \sum_{h \in [H]} y_{ih}(t)\frac{b_i}{B_i}\rceil-\sum_{j' =1}^{j-1} z_{ik_{j'}}(t),$
			\State ~~~~~~~~~~~~~~~~~~~~$\sum_{h \in [H]} y_{ih}(t)-\sum_{j' =1}^{j-1} z_{ik_{j'}}(t)\Big\}$
		\EndFor\label{paradeployend}
	
		\If{$\sum_{k \in [K]} z_{ik}(t) < \frac{b_i}{B_i} \sum_{h \in [H]} y_{ih}(t)$}/*not enough parameter servers can be deployed*/\label{checkparastart}
			\State Return $cost\_t=+\infty, \mathbf{y}, \mathbf{z}$
		\EndIf\label{checkparaend}
		
		\State $cost\_t=\sum_{h\in[H]}\sum_{r\in[R]} p_h^r(t)w_i^r y_{ih}(t) + \sum_{k\in[K]}\sum_{r\in[R]} q_k^r(t)s_i^r z_{ik}(t)$\label{costt_compute}
		\State Return $cost\_t, \mathbf{y}(t), \mathbf{z}(t)$\label{returncost}
	\EndFunction\label{cost_t_end}
	
\end{algorithmic}
}
\end{algorithm}

In Alg.~\ref{algo2}, we enumerate job completion time slot $\hat{t}_i$ (line \ref{lineenumtime}) and find the optimal schedule with each $\hat{t}_i$ by calling function {\bf DP\_COST}
(line \ref{lineilp}). Then we compare the payoffs achieved by schedules at different completion times and decide the best schedule achieving the highest payoff (lines \ref{mu_il}-\ref{linesaveend}). 

Lines \ref{DP_cost_begin}-\ref{DP_cost_end} implement a dynamic programming function:

\vspace{-5mm}
{\small
\begin{align}
	cost(\hat{t}_i, E_iN_i) =\min_{d\in[0,E_iN_i]} cost\_t({\hat{t}_i}, d)+cost(\hat{t}_i-1, E_iN_i-d)\nonumber
\end{align}
}
\vspace{-5mm}

\noindent We enumerate training workload $d$ to be finished in time slot $\hat{t}_i$ from $0$ to $E_iN_i$ (lines \ref{calcostbegin}-\ref{costt}), and let the rest workload $E_iN_i-d$ be carried out in $[a_i,\hat{t}_i-1]$ (line \ref{dpcall}). We compare the resulting costs (value of objective function (\ref{oracleilp})) and identify the schedule achieving the smallest cost (lines \ref{compcostbegin}-\ref{compcostend}). Finding the best schedule for workload $E_iN_i-d$ in $[a_i,\hat{t}_i-1]$ is the same as finding the best schedule to carry out workload $E_iN_i$ in $[a_i,\hat{t}_i]$ except for at a smaller scale, and hence the function calls itself in line \ref{dpcall} (a.k.a.~dynamic programming). Note that we always store the results of $COST\_t(t, d)$ and $DP\_COST(\mathcal{T}_i, \mathcal{D}_i)$ computed at different $\hat{t}_i$'s, to avoid re-computing the same subproblem in later iterations.

{\bf COST\_t} in lines \ref{cost_t_begin}-\ref{cost_t_end} computes the optimal worker and parameter server deployment to fulfil workload $d$ in time slot $t$. We sort servers for worker deployment in non-decreasing order of overall resource price $\sum_{r\in[R]}p_h^r(t)w_i^r$ (line \ref{sortworker}), and maximally deploy workers starting from the cheapest server, respecting capacity constraint (\ref{oracleconstraint4}) and upper bound $N_i$ on the number of workers in (\ref{constraint2}), to fulfil workload $d$ (lines \ref{workload}-\ref{workerdeployend}). 
Parameter servers are deployed in a similar greedy fashion.
The total number of parameter servers guarantees sufficient bandwidth to serve workers (constraint (\ref{constraint8})) but not over-provisioning (constraint (\ref{constraint9})), subject to capacity constraint (\ref{oracleconstraint5}) (lines \ref{paradeploystart}-\ref{paradeployend}). 
If not enough workers or parameter servers can be deployed, fulfilling workload $d$ in $t$ is infeasible 
(lines \ref{checkworkerstart}-\ref{checkworkerend}, \ref{checkparastart}-\ref{checkparaend}); otherwise, we return total deployment cost in $t$ (value of objective function (\ref{greedyilp})) and the schedule. 

\subsection{Theoretical Analysis}\label{theoanalysis}

We next analyze our online algorithm in terms of correctness, time complexity, and competitive ratio. \opt{short}{All missing proofs can be found in our technical report \cite{techreport}.}

\begin{theorem}[Optimality of Subroutine]\label{thoptdual}
Alg.~\ref{algo2} produces an optimal solution of problem (\ref{dualoracle}), in which 
{\bf COST\_t} computes an optimal solution of problem (\ref{greedyilp}).
\end{theorem}

\opt{long}{The proof can be found in Appendix~\ref{proofoptdual}.}


\begin{theorem}[Correctness]\label{thcorrect}
{\em OASiS} in Alg.~\ref{algo1} (together with Alg.~\ref{algo2}) 
computes a feasible solution to problems (\ref{obj}) (\ref{outobj}) (\ref{dualobj}).
\end{theorem}

\opt{long}{The proof can be found in Appendix~\ref{proofcorrect}.}

\noindent Though our online algorithm involves a dynamic programming approach, we prove its polynomial time complexity as follows.

\begin{theorem}[Polynomial Running Time]\label{thpolytime}
{\em OASiS} in Alg.~\ref{algo1} (together with Alg.~\ref{algo2}) 
runs in polynomial time to decide job admission and schedule upon arrival of each job $i$, with time complexity $O(TN_iE_i(H+K) + TN_i^2E_i^2)$.
\end{theorem}

\opt{long}{The proof can be found in Appendix~\ref{proofpolytime}.}

\noindent The competitive ratio of our online algorithm is the worst-case upper bound of the ratio of the overall utility of admitted jobs derived by the offline optimal solution of (\ref{obj}) to the total utility of admitted jobs achieved by Alg.~\ref{algo1} in the overall system span. 
\begin{theorem}[Competitive Ratio]\label{thratio}
{\em OASiS} in Alg.~\ref{algo1} is $2\alpha$-competitive, where $\alpha=\max_{r\in[R]}(1,\ln \frac{U_1^r}{L_1},\ln \frac{U_2^r}{L_2})$ and $U_1^r$, $U_2^r$, $L_1$ and $L_2$ are defined in (\ref{U_1})-(\ref{L_2}).
\end{theorem}

\opt{long}{The proof can be found in Appendix~\ref{proofratio}.}

Theorem \ref{thratio} tells that the larger the ratio of the largest utility to the lowest utility that the jobs can achieve is, the worse the ratio is. In this case, if {\em OASiS} makes a wrong decision, the gap from the offline optimum is larger. If the timespan $T$ or the total amount of resources is larger, the ratio is also worse, as there is more room for the offline algorithm to improve. 

\section{Performance Evaluation}\label{evaluation}

We next evaluate {\em OASiS} by simulation studies and testbed experiments based on a prototype system implementation. 

\subsection{Simulation Studies}

\noindent{\bf Settings.}
We simulate an ML system running for $T=100$-$300$ time slots, with $H=50$ servers to host workers (server resource capacities set according to Amazon EC2 C4 instances) and $K=50$ servers to deploy parameter servers (resource capacities following EC2 GPU instances P2 and G3 randomly \cite{amazongpuserver}). 
Server bandwidth is set within $[20,50]$Gbps. Following similar settings in \cite{sun2017towards}\cite{li2014scaling}\cite{chilimbi2014project}, we set resource demands of each worker as follows: $0$ to $4$ GPUs, $1$ to $10$ vCPUs, $2$ to $32$GB memory, $5$ to $10$GB storage, and bandwidth of $100$Mbps to $5$Gbps ($b_i$). Resource demands of each parameter server are: $1$ to $10$ vCPUs, $2$ to $32$GB memory, $5$ to $10$GB storage, and bandwidth of $5$Gbps to $20$Gbps ($B_i$). 
We set job arrival pattern according to the Google cluster data \cite{reiss2012heterogeneity}, but may vary job arrival rates. 
For different jobs, $E_i$ is set within $[50,200]$, $N_i$ is in $[5,100]$, $M_i$ is in $[10,100]$, $\tau_i$ is in $[0.001,0.1]$ time slots, and $e_i$ is within $[30,575]$MB \cite{iandola2016firecaffe}. 
We use 
a sigmoid utility function \cite{huang2015need}, $f_i(t-a_i)=\frac{\gamma_1}{1+e^{\gamma_2(t-a_i-\gamma_3)}}$, where $\gamma_1$ is priority of job $i$ in $[1,100]$, $\gamma_2$ is a decay factor, and $\gamma_3$ is the target completion time of job $i$ set in $[1,15]$. We set $\gamma_2=0$ for time-insensitive jobs (constant utility), $\gamma_2$ in $[0.01,1]$ to represent time-sensitive jobs and $\gamma_2$ in $[4,6]$ for time-critical jobs. By default, $10\%$, $55\%$ and $35\%$ jobs are time-insensitive, -sensitive, and -critical, respectively, in our experiments.

\noindent {\bf Schemes for comparison.} We compare {\em OASiS} with four representative job scheduling policies in existing cloud platforms.
(i) FIFO: default scheduler in Hadoop and Spark \cite{zaharia2010spark}
; jobs are admitted and run in order of their arrivals, with fixed numbers of workers/parameter servers. 
(ii) Dominant Resource Fairness Scheduling (DRF): default scheduler in YARN \cite{vavilapalli2013apache} and Mesos \cite{hindman2011mesos}; jobs are all admitted and numbers of workers/parameter servers are computed to achieve max-min fairness in dominant resources upon job arrival and job completion \cite{ghodsi2011dominant}. 
(iii) Risk-Reward Heuristic (RRH) \cite{irwin2004balancing}: a job is admitted if its utility minus a delay cost incurred by its admission is larger than a threshold; upon job arrival or completion, unfinished jobs either continue running (always with same worker/parameter server numbers once running) or pause, decided by job's future utility gain minus cost.
(iv) Dorm \cite{sun2017towards}: Jobs are admitted; upon job arrival or completion, numbers and placement of workers/parameter servers of unfinished jobs are recomputed by an MILP resource utilization maximization problem, subject to fairness and adjustment overhead constraints. 
In (i)-(iii), 
we place workers and parameter servers on available servers in a round-robin fashion. For FIFO and RRH, the number of workers (parameter servers) is fixed to a number within $[1, 30]$. 

\begin{figure}[!t]
\begin{minipage}[t]{0.48\linewidth}
\centering
  \includegraphics[width=1.8in]{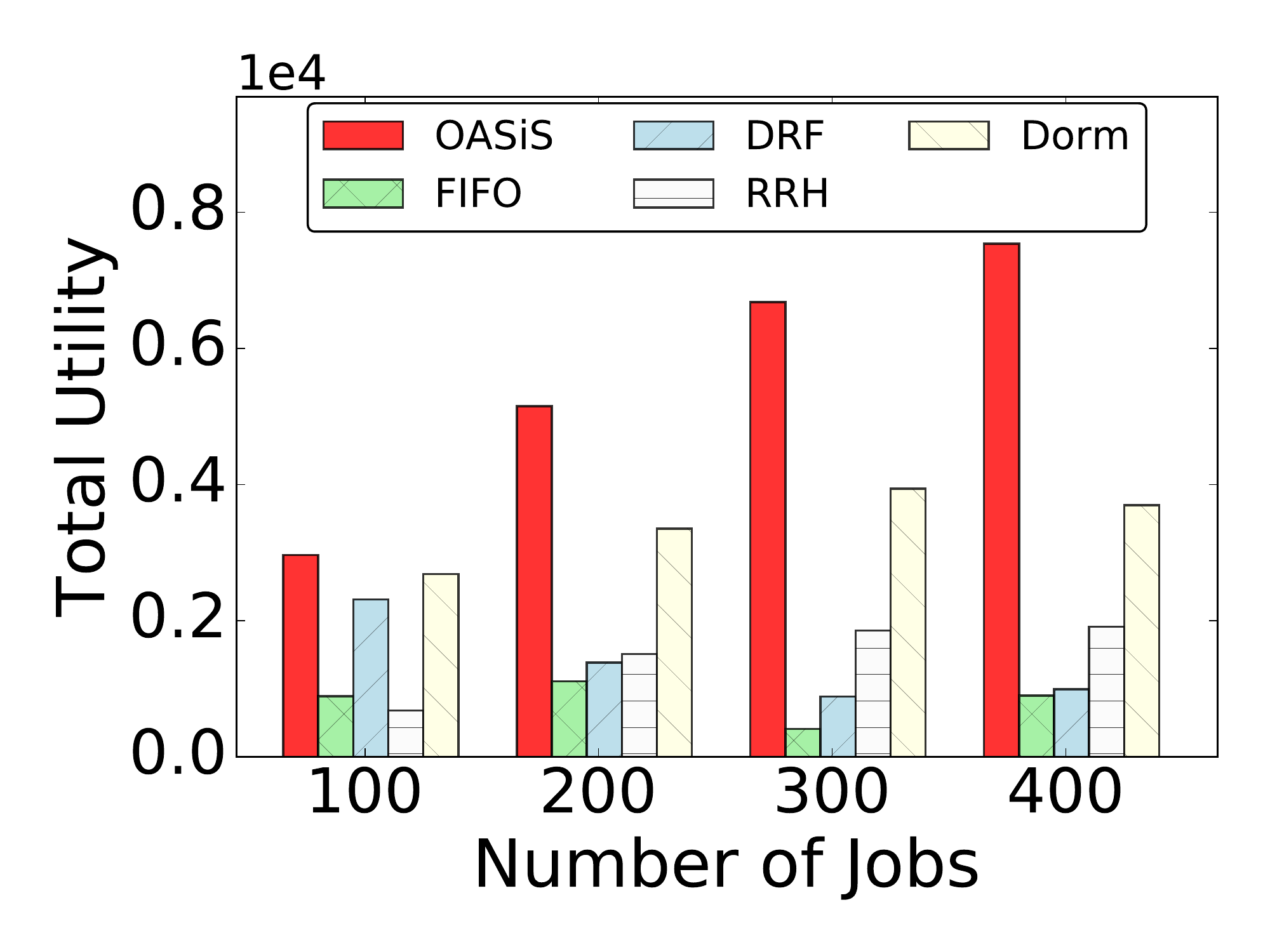}
  \caption{Total job utility}
  \label{figutility}
\end{minipage}
\hfill
\begin{minipage}[t]{0.48\linewidth}
\centering
\includegraphics[width=1.8in]{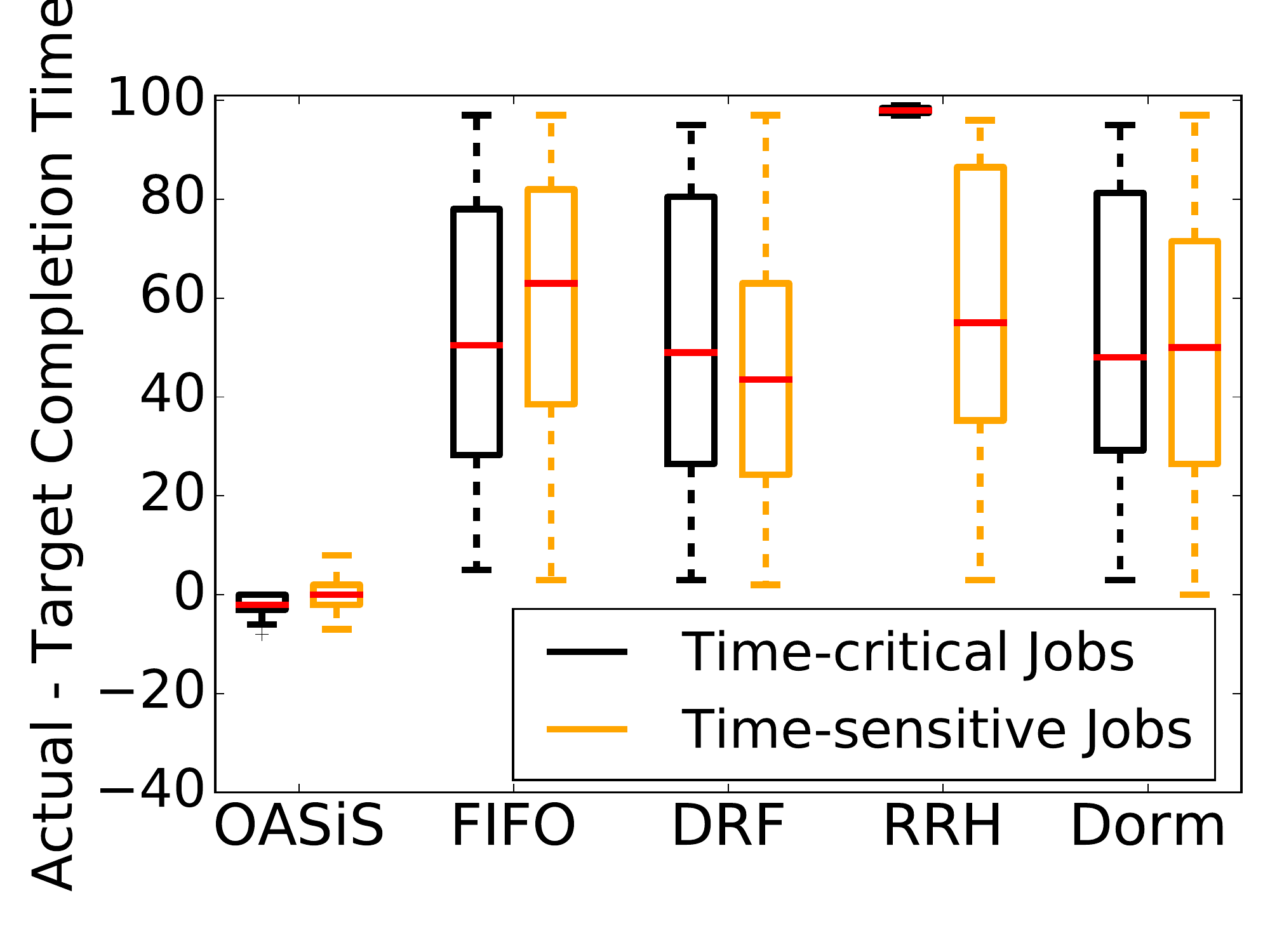}
\caption{Completion timeliness}
\label{figgap}
\end{minipage}
\hfill
\end{figure}

\begin{figure}[!t]
\begin{minipage}[t]{0.48\linewidth}
\centering
\includegraphics[width=1.8in]{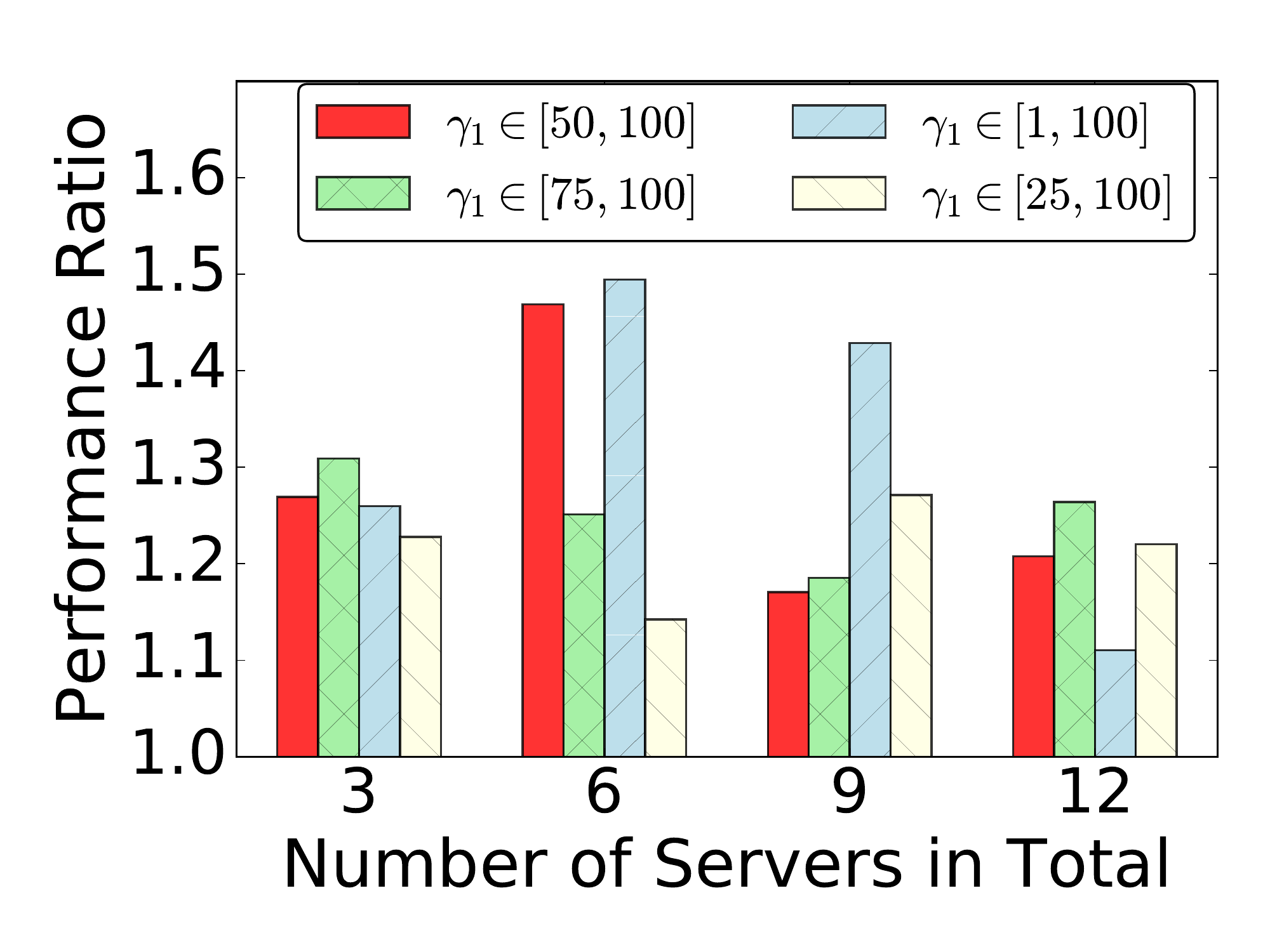}
\caption{Performance ratio}
\label{figratio}
\end{minipage}
\hfill
\begin{minipage}[t]{0.48\linewidth}
\centering
  \includegraphics[width=1.8in]{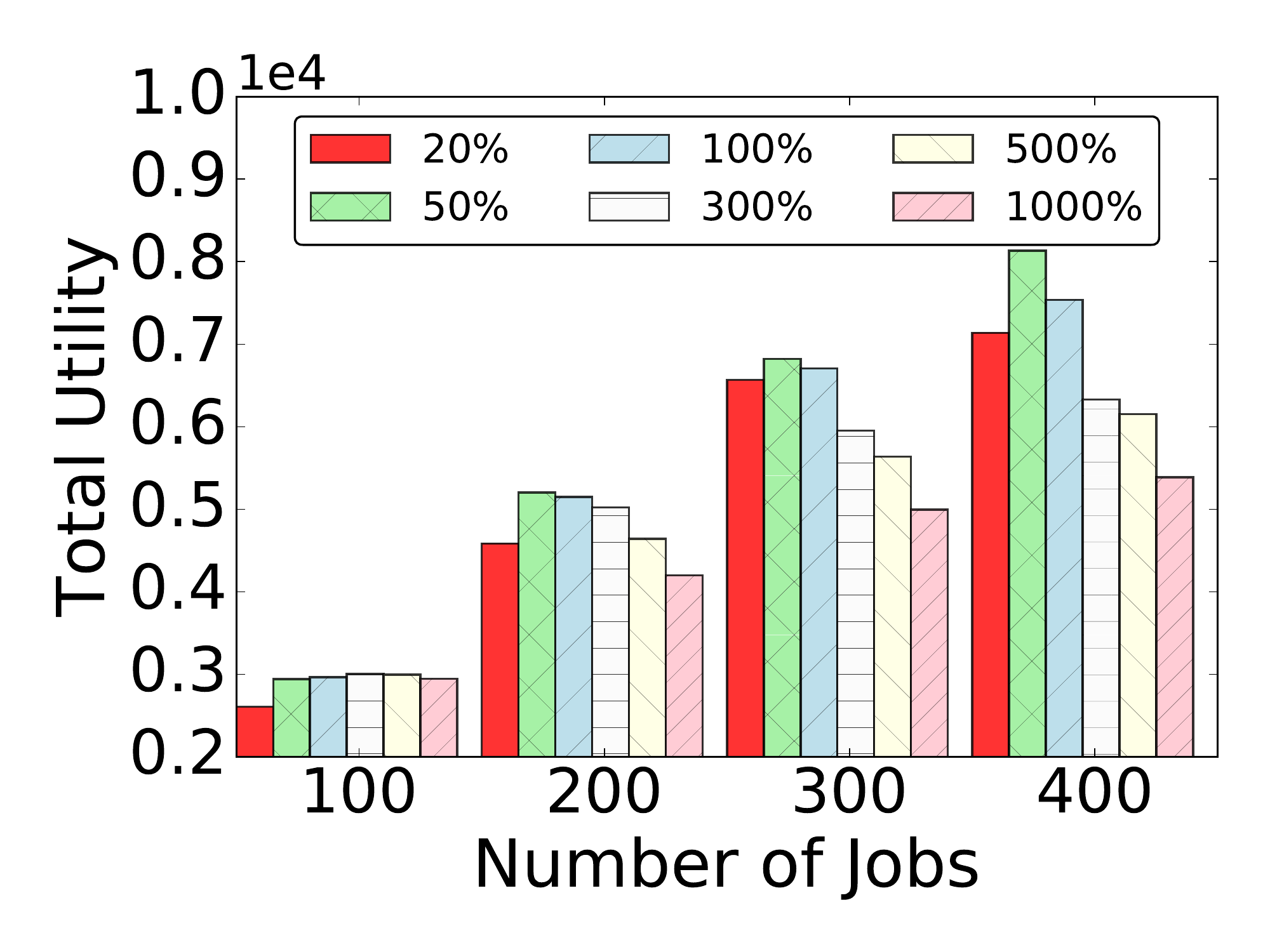}
  \caption{Total job utility under inaccurate $\frac{U_1^r}{L_1}$, $\frac{U_2^r}{L_2}$}
  \label{figchg}
\end{minipage}
\hfill
\end{figure}
%

\noindent{\bf Results.}  
Fig.~\ref{figutility} presents the total utility achieved by different schemes, where $T=300$. 
 {\em OASiS} performs the best, especially when the number of jobs in the fixed timespan is larger (resources are more scarce).

Fig.~\ref{figgap} shows how well the target completion time is met when $100$ time-sensitive and $100$ time-critical jobs are run in $T=100$. The actual completion time minus target completion time ($\gamma_3$ in the sigmoid utility function) achieved with {\em OASiS} is the closest to zero for both types of jobs, with the smallest variance. 
Among the other schemes, only RRH is job utility (completion time) aware, but its resource utilization is not as efficient so does not perform well either. 

Fig.~\ref{figratio} shows the performance ratio of {\em OASiS}, computed by dividing the total job utility of the offline optimal solution by the total job utility achieved by {\em OASiS}. Due to the time complexity of solving (\ref{obj}) exactly for the offline optimum, the number of jobs is limited to $10$.\footnote{It takes 2 days to compute the optimal offline solution with $10$ jobs, while {\em OASiS} runs for less than $1$ second to produce the best schedule for each job in the case of $100$ time slots and $80$ worker/parameter servers.} We set $T=10$, vary the number of servers (proportionally divided to host workers and parameter servers), and also vary the range of job priorities ($\gamma_1$ in the sigmoid function), such that $\max_{r\in[R]}(\frac{U_1^r}{L_1}, \frac{U_2^r}{L_2})$ increases from left to right at each fixed number of servers in the figure. 
We observe a ratio around $1.1$ to $1.5$, showing the good performance of our online algorithm. There is no clear trend of increase or decrease of the ratio with more resources and larger $\max_{r\in[R]}(\frac{U_1^r}{L_1}, \frac{U_2^r}{L_2})$ -- the factors influencing the worst-case competitive ratio in Theorem \ref{thratio} (note our simulation scenario may not be the worst case). 

In Fig.~\ref{figchg}, we use estimated values of $\frac{U_1^r}{L_1}$ and $\frac{U_2^r}{L_2}$ as input to {\em OASiS}, at different percentages of their actual values ($T=300$). 
We observe that an underestimation leads to higher total utility than overestimation when resources are scarce, as it prevents abrupt price rise which may filter out jobs that should be accepted. These results directly reflect impact of using inaccurate estimations of $\frac{U_1^r}{L_1}$ and $\frac{U_2^r}{L_2}$ on performance ratio of {\em OASiS}. 

\subsection{Testbed Experiments}
%
%
%

\noindent{\bf Prototype implementation.} 
We implement a distributed ML system based on MXNet \cite{chen2016mxnet} with Kubernetes 1.6 \cite{Kubernetes_blog}. MXNet is modified to support dynamic adjustment of worker/parameter server numbers. {\em OASiS} and other 4 scheduling schemes for comparison are implemented as custom schedulers 
to replace the default one in Kubernetes, respectively. 
The scheduler constantly queries ongoing jobs and available system resources, and posts scheduling decisions via the Kubernetes API server. Each worker or parameter server is implemented on a Docker container with $0$ to $1$ GPU, $1$ to $5$ CPU cores, $2$ to $10$GB memory, and $1$ to $3$Gbps bandwidth. 
\opt{long}{An illustration of the testbed architecture is given in Fig.~\ref{figtestbed}. }
We deploy our system on 9 servers: 2 with 8 CPU cores,  32GB RAM, 1.5TB storage each host parameter servers, and 7 with 32 CPU cores, 80GB RAM, 600GB storage each host workers (there are 4 GPUs deployed on 4 servers). All servers are equipped with a dual-port 1GbE NIC and a dual-port 10GbE NIC. 
All data are stored in HDFS \cite{hdfs}, with chunk size 2MB.


\noindent{\bf Experimental setup.}
We run 6 kinds of model training jobs, \ie, AlexNet \cite{krizhevsky2012imagenet}, ResNet-50,101,152 \cite{he2016deep}, VGG-11 \cite{simonyan2015very}, and Inception-BN \cite{ioffe2015batch}, on ImageNet ILSVRC2012 \cite{deng2009imagenet} dataset (we use 200 images (20.3MB)). Each experiment runs for $10$ time slots and each time slot is $20$ minutes long. 12 jobs arrive 
in the first 9 time slots and each job runs for $40$ minutes to $2$ hours.  
Each data chunk contains 20 or 30 images, each mini-batch contains 10 images, and the number of epochs is in $[4,30]$. Job utilities are similar to simulation. 

\opt{long}{
\begin{figure}[!t]
\centering
  \includegraphics[width=1.8in]{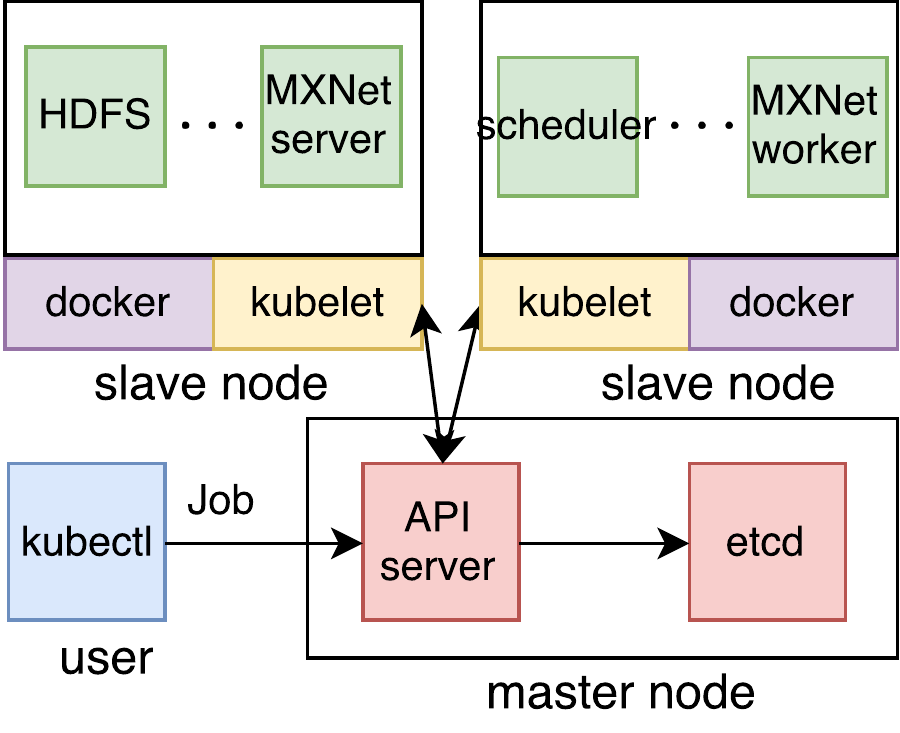}
  \caption{Testbed}
  \label{figtestbed}
\end{figure}
}

\begin{figure}[!t]
\begin{minipage}[t]{0.48\linewidth}
\centering
  \includegraphics[width=1.8in]{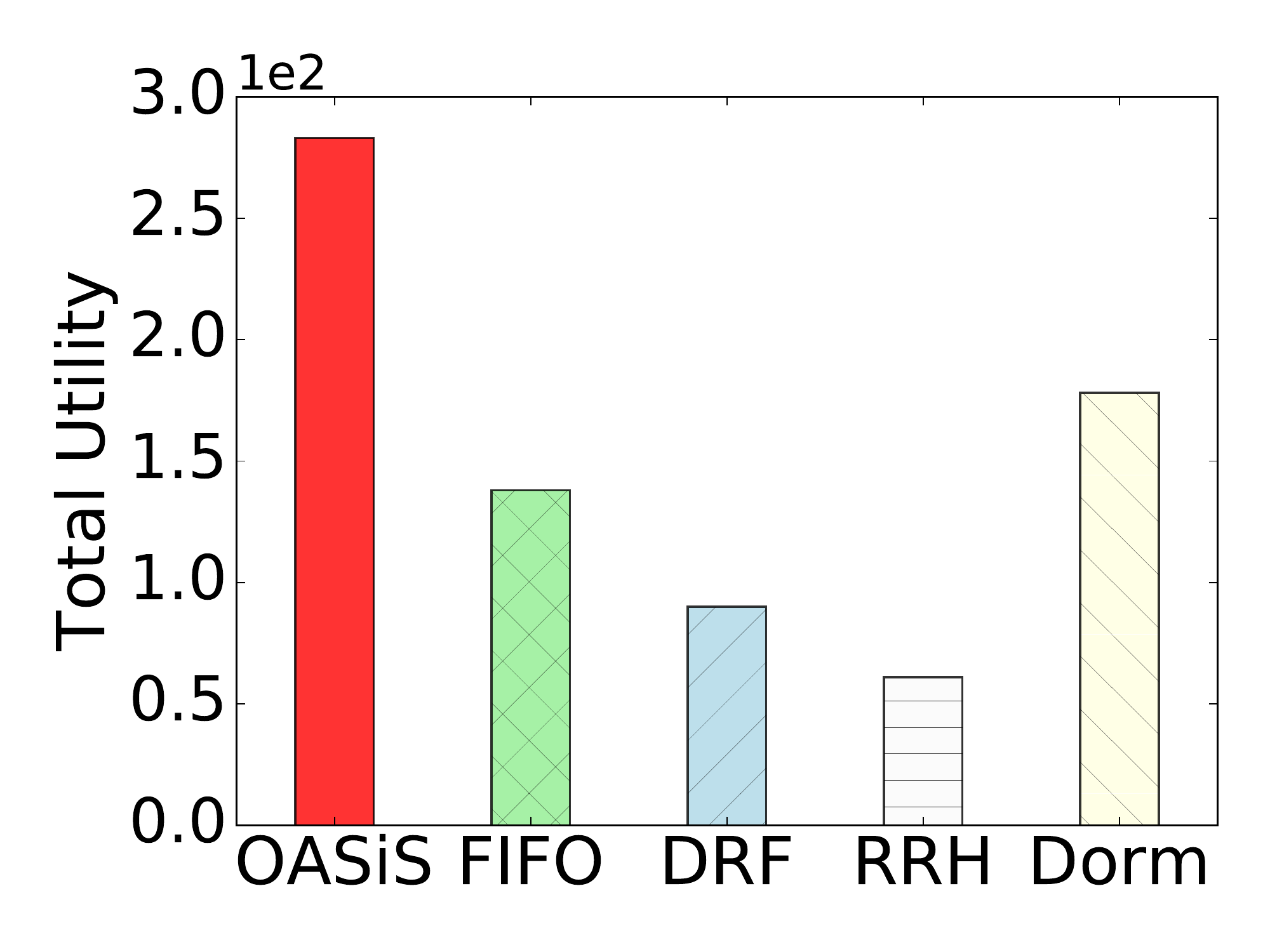}
  \caption{Total job utility}
  \label{figutilityexp}
\end{minipage}
\hfill
\begin{minipage}[t]{0.48\linewidth}
\centering
\includegraphics[width=1.8in]{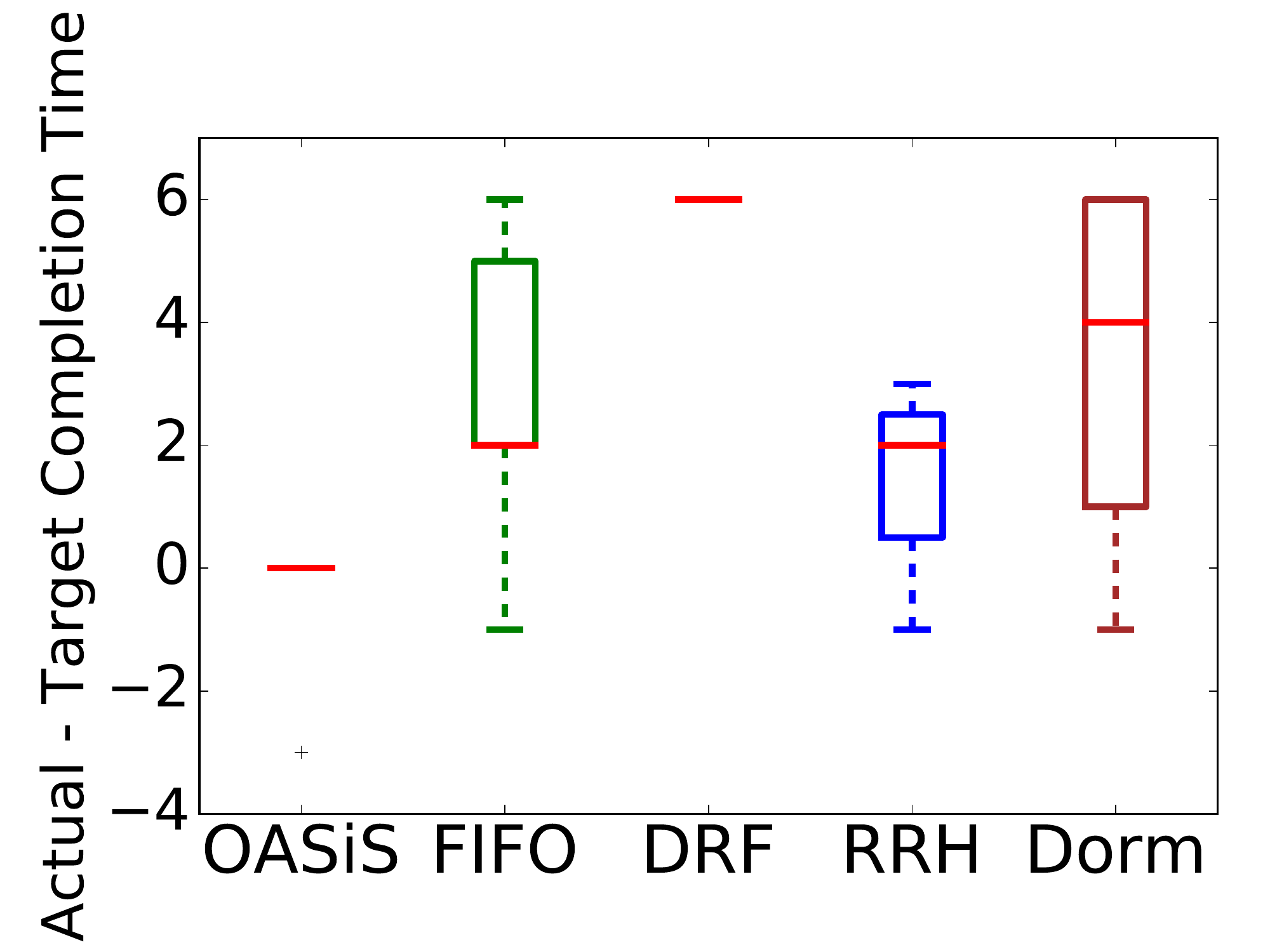}
\caption{Completion timeliness}
\label{figgapexp}
\end{minipage}
\hfill
\end{figure}

\noindent{\bf Experimental results.}
We plot the total utility in Fig.~\ref{figutilityexp} and the actual completion time minus target completion time of all jobs in Fig. \ref{figgapexp}. 
Compared to Fig.~\ref{figutility} and Fig.~\ref{figgap}, the comparison results are similar. 
With the small number of jobs that we can run on our small testbed, the difference between {\em OASiS} and others may not be as apparent as that in a large system (as shown by our larger scale simulations). We are confident that the advantage of our algorithm will be more obvious when experimenting on a large testbed. 

\section{Conclusion}\label{conclusion}
This paper proposes {\em OASiS}, an online algorithm for admission and scheduling of asynchronous training jobs in an ML cluster. {\em OASiS} computes the best schedule to run each job, using a varying number of workers and parameter servers over time for best resource utilization and training expedition, while admitting jobs judiciously based on carefully set 
resource prices, for long-term utility maximization. Our theoretical analysis shows polynomial running time and a good competitive ratio of {\em OASiS}. Simulation and experiments on a prototype system show that {\em OASiS} outperforms common schedulers in real-world cloud systems.

\bibliographystyle{IEEEtran}
\bibliography{IEEEabrv,reference}

\opt{long}{\begin{appendices}
\section{The Property of Price Functions (\ref{marginalcosts})}\label{proofmaginalcosts}

Note that $U_1^r$, $U_2^r,\forall r \in[R]$, $L_1$ and $L_2$ in (\ref{U_1})-(\ref{L_2}) are equivalent to
\begin{align}
& U_1^r=\max_{i\in[I],l\in\mathcal{L}_i} \frac{f_i(t_{il}-a_i)}{w_i^r},\forall r\in[R]\nonumber\\
& U_2^r=\max_{i\in[I],l\in\mathcal{L}_i} \frac{f_i(t_{il}-a_i)}{s_i^r},\forall r\in[R]\nonumber\\
& L_1=\frac{1}{4 \eta_1}\min_{i\in[I],l\in\mathcal{L}_i} \frac{f_i(t_{il}-a_i)}{\sum_{r\in[R]} \lceil E_i N_i M_i (\tau_i + 2e_i/b_i)\rceil w_i^r}\nonumber\\
& L_2=\frac{1}{4 \eta_2}\min_{i\in[I],l\in\mathcal{L}_i} \frac{f_i(t_{il}-a_i)}{\sum_{r\in[R]} \lceil E_i N_i M_i (\tau_i + 2e_i/b_i)\rceil s_i^r}\nonumber
\end{align}

When there is no resource usage of all types of resources on all servers at time $t$, {\em i.e.}, $g_h^r(t)=0,\forall h\in[H],r\in[R]$, $v_k^r(t)=0,\forall k\in[K],r\in[R]$ we have $p_h^r(t)=L_1,q_k^r(t)=L_2, \forall h\in[H], k\in[K],r\in[R]$. In (\ref{outdualconstraint1}), we have

\begin{align*}
& \mu_i \geq f_i(t_{il}-a_i) - \sum_{r\in[R]} \sum_{t \in l} \sum_{h \in (t,l)} L_1 w_i^r y_{ih}^{l}(t)\\
& - \sum_{r\in[R]} \sum_{t \in l} \sum_{k \in (t,l)} L_2 s_i^r z_{ik}^{l}(t)\\
& \geq f_i(t_{il}-a_i) - \sum_{r\in[R]} \sum_{t \in l} \sum_{h \in (t,l)}\nonumber\\
& \frac{1}{4 \eta_1}\frac{f_i(t_{il}-a_i)}{\sum_{r\in[R]} \lceil E_i N_i M_i (\tau_i + 2e_i/b_i)\rceil w_i^r} w_i^r y_{ih}^{l}(t)\nonumber\\
& - \sum_{r\in[R]} \sum_{t \in l} \sum_{k \in (t,l)} \frac{1}{4 \eta_2} \frac{f_i(t_{il}-a_i)}{\sum_{r\in[R]} \lceil E_i N_i M_i (\tau_i + 2e_i/b_i)\rceil s_i^r} s_i^r z_{ik}^{l}(t)\\
& \geq f_i(t_{il}-a_i) - \sum_{t \in l} \sum_{h \in (t,l)} \frac{1}{4 \eta_1} \frac{f_i(t_{il}-a_i)}{\lceil E_i N_i M_i (\tau_i + 2e_i/b_i)\rceil}y_{ih}^{l}(t)\\
& - \sum_{t \in l} \sum_{k \in (t,l)} \frac{1}{4 \eta_2} \frac{f_i(t_{il}-a_i)}{\lceil E_i N_i M_i (\tau_i + 2e_i/b_i)\rceil}z_{ik}^{l}(t)\\
& =(1-\frac{1}{4\eta_1}-\frac{1}{4\eta_2})f_i(t_{il}-a_i)
>0
\end{align*}

Note that $\frac{1}{\eta_1}\leq 1$ and $\frac{1}{\eta_2}\leq 1$.

Then any job can be scheduled in this condition.

Besides, given a schedule $l$ for job $i$, if there is no resource usage for all $t\in l, h\in(t,l),k\in(t,l)$, job $i$ must be accepted due to the same reason.


Similarly, when type-$r$ resource on server $h$ is exhausted at time slot $t$, {\em i.e.}, $\exists t\in[T], h\in[H], r\in[R], g_h^r(t)=c_h^r$, we have $p_h^r(t)=U_1^r$. In (\ref{outdualconstraint1}), if the schedule $l$ includes the time slot $t$ and server $h$, and if we plan to accept job $i$ with schedule $l$, we have

\begin{align*}
& \mu_i = f_i(t_{il}-a_i) - U_1^r w_i^r y_{ih}^{l}(t)-\sum_{r'\in[R]} \sum_{t \in l} \sum_{h' \in (t,l)} p_{h'}^{r'}(t) w_i^{r'}y_{ih'}^{l}(t)\\
& - \sum_{r\in[R]} \sum_{t \in l} \sum_{k \in (t,l)} q_{k}^{r}(t) s_i^{r} z_{ik}^{l}(t)\\
& \leq f_i(t_{il}-a_i) - \frac{f_i(t_{il}-a_i)}{w_i^r} w_i^r y_{ih}^{l}(t) - \sum_{r'\in[R]} \sum_{t \in l} \sum_{h' \in (t,l)} p_{h'}^{r'}(t)\\
& w_i^{r'} y_{ih'}^{l}(t) - \sum_{r\in[R]} \sum_{t \in l} \sum_{k \in (t,l)} q_{k}^{r}(t) s_i^{r} z_{ik}^{l}(t)\\
& \leq f_i(t_{il}-a_i) - f_i(t_{il}-a_i) y_{ih}^{l}(t) - \sum_{r'\in[R]} \sum_{t \in l} \sum_{h' \in (t,l)} p_{h'}^{r'}(t) w_i^{r'}\\
& y_{ih'}^{l}(t) - \sum_{r\in[R]} \sum_{t \in l} \sum_{k \in (t,l)} q_{k}^{r}(t) s_i^{r} z_{ik}^{l}(t)\\
& \leq 0
\end{align*}
where $r'$, $h'$ are other resources and servers such that the resource of type-$r'$ on server $h'$ is not full at time $t$.

Then for any job $i$, its payoff $\mu_i\leq0$, {\em OASiS} must reject that job. No job can be scheduled at time $t$ if one or more of the resources is used up.

Similarly, when type-$r$ resource on server $k$ is exhausted at time slot $t$, {\em i.e.}, $\exists t\in[T], k\in[K], r\in[R], v_k^r(t)=c_k^r$, we have $q_k^r(t)=U_2^r$ and $\mu_i\leq 0$.

\section{Proof of Theorem~\ref{thoptdual}}\label{proofoptdual}
\begin{IEEEproof}
\subsection{Optimality of Function {\bf COST\_t} and {\bf DP\_COST}}
We prove the optimality of function {\bf COST\_t} and {\bf DP\_COST} for each given $\hat{t}_i$ first.

We firstly prove that calculating the minimum cost to fulfill training workload $d$ at time $t$ for one job, {\em i.e.}, $cost\_t(t,d)$ is optimal using greedy algorithm. Denote the schedule as $S_{t,d}$.

Note that $\lceil d M_i (\tau_i + 2e_i/b_i)\rceil$ is the minimum number of workers fulfill workload of $d$. Given the value of $\sum_{h\in[H]} y_{ih}(t)$, $\lceil \frac{b_i}{B_i} \sum_{h\in[H]} y_{ih}(t)\rceil$ is the minimum number of parameter servers to satisfy constraints (\ref{constraint8})(\ref{constraint9}). If there is another deployment $S_{t,d}'$ whose cost is smaller than $cost\_t(t,d)$, denote the worker deployment and parameter server deployment as $y_{ih}'(t), \forall h\in[H]$ and $z_{ik}'(t),\forall k\in[K]$, respectively. There are two cases of this worker deployment.

\begin{enumerate}
\item If $\sum_{h\in[H]}y_{ih}'(t)=\sum_{h\in[H]} y_{ih}(t)$, then in $S_{t,d}'$ there exists some $y_{ih}'(t)$ such that $y_{ih}'(t)\neq y_{ih}(t)$. As the total number of workers keeps the same, this is equivalent to move some workers from their servers in $S_{t,d}$ to other servers in $S_{t,d}'$. Since we deploy workers on the servers from the one with the lowest resource cost per worker to the one with the highest resource cost per worker greedily, moving workers from their deployed servers $S_{t,d}$ to other servers can only increase the total cost.
\item If $\sum_{h\in[H]}y_{ih}'(t)>\sum_{h\in[H]} y_{ih}(t)$, then we need to add more workers, which leads to the increase of total cost. From the first case, we know that changing the worker placement in $S_{t,d}$ while keeping the total number of workers the same can only increase the total cost. So in this case, no matter we add some new workers to $S_{t,d}$ or change the placement after adding new workers, the total cost will be increased.
\end{enumerate}

For the deployment of parameter servers, as $\sum_{h\in[H]}y_{ih}'(t)\geq\sum_{h\in[H]} y_{ih}(t)$, we have $\sum_{k\in[K]}z_{ik}'(t)\geq\sum_{k\in[K]} z_{ik}(t)$. There are also two cases.

\begin{enumerate}
\item If $\sum_{k\in[K]}z_{ik}'(t)=\sum_{k\in[K]} z_{ik}(t)$, then in $S_{t,d}'$ there exists some $z_{ik}'(t)$ such that $z_{ik}'(t)\neq z_{ik}(t)$. As the total number of parameter servers keeps the same, this is equivalent to move some parameter servers from their servers in $S_{t,d}$ to other servers in $S_{t,d}'$. Similarly, moving parameter servers from their deployed servers $S_{t,d}$ to other servers can only increase the total cost.
\item If $\sum_{k\in[K]}z_{ik}'(t)>\sum_{k\in[K]} z_{ik}(t)$, then we need to add more parameter servers, which leads to the increase of total cost due to similar reason.
\end{enumerate}

From the analysis above, we know that computing $cost\_t(t,d),\forall t\in[a_i,\mathcal{T}_i],0\leq d\leq N_iE_i$ is optimal.

Next we prove the optimality of computing $cost(t,d)$ by the induction on $t=a_i,\dots,\mathcal{T}_i$.

Note that $cost(a_i,d)=cost\_t(a_i,d),\forall 0\leq d\leq N_iE_i$, so $cost(a_i,d),\forall 0\leq d\leq N_iE_i$ is the minimum cost to fulfill training workload $d$ at time $a_i$.

The induction hypothesis shows that $cost(t,d)$ is the minimum cost to fulfill training workload $d$, at time $t$, for all $0\leq d\leq N_iE_i$. According to the dynamic programming function, we have

\begin{align*}
& cost(t+1,d) = \min_{0\leq d' \leq d} \{cost\_t(t+1,d') + cost(t,d - d')\},\\
& \forall 0\leq d\leq N_iE_i
\end{align*}

Based on the induction hypothesis, $cost(t+1,d)$ is also the minimum cost to fulfill training workload $d$ at time $t+1$.

Then we can conclude $cost(t,d)$ is the minimum cost to fulfill workload of $d$ in time $[a_i, t]$, for all $t\in[a_i,\mathcal{T}_i],0\leq d\leq N_iE_i$.

\subsection{Optimality of Alg.~\ref{algo2}}

Note that in Alg.~\ref{algo2}, $\hat{t}_i$ is the deadline instead of the job completion time slot of job $i$. We show that after the enumeration from $a_i$ to $T$, the optimal $\hat{t}_i$ is tight, {\em i.e.}, $\sum_{h\in[H]} y_{ih}(\hat{t}_i)>0,\sum_{h\in[H]} y_{ih}(\hat{t}_i+1)=0$.

Assume the optimal solution, {\em i.e.}, the best schedule $l^*$ finishes the job $i$ at $\hat{t}^*_i$. Then when we enumerate $t=\hat{t}_i^*$, we must derive the corresponding schedule $l=l^*$.

We enumerate $\hat{t}_i$ from $a_i$ to $T$ and only update the best schedule when the payoff of the new schedule is smaller than the current one. Note that utility function $f_i(t-a_i)$ is non-increasing, if there exist another deadline $\hat{t}_i'>\hat{t}_i^*$ and with the same schedule such that $l'=l^*$, the objective value (\ref{dualoracle}) can only be equal or smaller than that of $l^*$. So we will stick to the optimal solution $l^*$ once we find it.

Then Alg.~\ref{algo2} produces the optimal solution of (\ref{dualoracle}).

\end{IEEEproof}

\section{Proof of Theorem~\ref{thcorrect}}\label{proofcorrect}
\begin{IEEEproof}
Alg.~\ref{algo2} and (\ref{oracleilp}) guarantee that the schedule $l$ for job $i$ satisfies constraints (\ref{constraint1})-(\ref{constraint8}),  (\ref{constraint10})-(\ref{constraint13}), (\ref{outconstraint1})-(\ref{outconstraint2}). In function $COST\_t(t, d)$, as the assignment of $z_{ik}(t)$ stops when $\sum_{k\in[K]}z_{ik}(t)=\lceil \frac{b_i}{B_i} \sum_{h \in [H]} y_{ih}(t) \rceil$, constraint (\ref{constraint9}) must hold. Constraint (\ref{outconstraint3}) holds as we only produce one schedule for each job in Alg.~\ref{algo2}. For the dual problem (\ref{dualobj}), Alg.~\ref{algo1} let $\mu_i$ be $0$ if $f_i(t_{il}-a_i) \leq \sum_{r\in[R]} \sum_{t \in l} \sum_{h \in (t,l)} p_h^r(t) w_i^r y_{ih}^{l}(t) + \sum_{r\in[R]} \sum_{t \in l} \sum_{k \in (t,l)} q_k^r(t) s_i^r z_{ik}^{l}(t),\forall l\in\mathcal{L}_i$ and set $\mu_i =\max_{l\in\mathcal{L}_i} \Big(f_i(t_{il}-a_i) - \sum_{r\in[R]} \sum_{t \in l} \sum_{h \in (t,l)} p_h^r(t) w_i^r y_{ih}^{l}(t) - \sum_{r\in[R]} \sum_{t \in l} \sum_{k \in (t,l)} q_k^r(t) s_i^r z_{ik}^{l}(t)\Big)$ otherwise, which ensures the feasibility of dual problem (\ref{dualobj}).
\end{IEEEproof}

\section{Proof of Theorem~\ref{thpolytime}}\label{proofpolytime}

\begin{IEEEproof}
When executing the dual subroutine Alg.~\ref{algo2}, given end time $\hat{t}_i$, finding the best schedule that finishes job $i$ before $\hat{t}_i$ takes polynomial time. In function $COST\_t(t, d)$, sorting servers according to the cost per worker/parameter server under each time slot $t$ takes $O(H\log H)$ and $O(K\log K)$ time. To find the deployment $y_{ih}(t)$ and $z_{ik}(t)$, $\forall h\in[H],k\in[K]$, we need to loop all servers, which takes $O(H+K)$ time. So calculate all $cost\_t(t,d),\forall d\in[0,N_iE_i]$ under a given $t$ needs $O(H\log H + K\log K + N_iE_i(H+K))$ time. Note that the order of sorted servers can be saved once calculated and we do not need to sort them again under different workload $d$ at the same time slot. The number of states $(t,d)$ for each job $i$ in dynamic programming is $O(TN_iE_i)$, and the time complexity of executing dynamic programming function is $O(N_iE_i)$, as we already save all pre-calculated $cost\_t(t,d)$ and $cost(t,d)$ for all $t\in[T],d\in[0,N_iE_i]$. Then the time complexity of dynamic programming is $O(TN_i^2E_i^2)$. Enumerating all $\hat{t}_i$ from $a_i$ to $T$ in Alg.~\ref{algo2}, the time complexity to decide the best schedule for one job is $O(T(H\log H + K\log K) + TN_iE_i(H+K) + TN_i^2E_i^2)$. When updating variables in Alg.~\ref{algo1}, each statement is executed at most $O(TKV)$ or $O(THV)$ times. In conclusion, the time complexity for making a scheduling decision for job $i$ is $O(T(H\log H + K\log K) + TN_iE_i(H+K) + TN_i^2E_i^2 + TKV + THV)$, namely, $O(TN_iE_i(H+K) + TN_i^2E_i^2)$.
\end{IEEEproof}

\section{Proof of Theorem~\ref{thratio}}\label{proofratio}
\subsection{Preliminaries}
We denote $OPT$ as the optimal objective value of (\ref{obj}) and (\ref{outobj}). Let $P_i$ and $D_i$ be the objective value of primal problem (\ref{outobj}) and that of dual problem (\ref{dualobj}) respectively, returned by Alg.~\ref{algo1} after deciding the schedule of job $i$. Let $P_0$ and $D_0$ be the initial values of (\ref{outobj}) and (\ref{dualobj}). Note that $P_0=0$ and $D_0=\sum_{t\in[T]}\sum_{h\in[H]}\sum_{r\in[R]} P_h^r(0)c_h^r + \sum_{t\in[T]}\sum_{k\in[K]}\sum_{r\in[R]} Q_k^r(0)c_k^r$. Then $P_I$ and $D_I$ are the final primal and dual objective values returned by Alg.~\ref{algo1}.

\begin{lemma}\label{thpdratio}
If there exists a constant $\alpha\geq 1$ such that $P_i-P_{i-1}\geq \frac{1}{\alpha} (D_i - D_{i-1})$ for all jobs $i\in[I]$, and if $P_0=0$ and $D_0\leq\frac{1}{2}OPT$, then Alg.~\ref{algo1} is $2\alpha$-competitive in total job utility.
\end{lemma}
\begin{IEEEproof}
Note that $P_I$ is the summation of $P_i-P_{i-1}$ over all jobs $i\in[I]$, {\em i.e.}, $P_I=\sum_{i\in[I]}(P_i-P_{i-1})$. Similarly, $D_I-D_0=\sum_{i\in[I]}(D_i-D_{i-1})$. So we have
\begin{align*}
P_I=\sum_{i\in[I]}(P_i-P_{i-1})\geq \frac{1}{\alpha}\sum_{i\in[I]}(D_i-D_{i-1})=\frac{1}{\alpha}(D_I-D_0)
\end{align*}
According to weak duality \cite{boyd2004convex}, we have
\begin{align*}
D_I\geq OPT \geq P_I
\end{align*}
Then we can derive
\begin{align*}
& D_I-D_0 \geq \frac{1}{2} OPT\\
& P_I \geq \frac{1}{\alpha}(D_I-D_0) \geq \frac{1}{2\alpha} OPT
\end{align*}
So we can conclude that the competitive ratio is $2\alpha$.
\end{IEEEproof}

We introduce the relationship between the cost and resource consumption before and after processing one job. Let $p_h^{r,i}(t)$ ($q_k^{r,i}(t)$) be the unit cost of type-$r$ resource on server $h$ (server $k$) at time $t$ after handling job $i$. Let $g_h^{r,i}(t)$ ($v_k^{r,i}(t)$) be the amount of type-$r$ resource allocated to jobs on server $h$ (server $k$) at time $t$ after dealing with the job $i$.

\begin{definition}\label{defacr}
The allocation-cost relationship for Alg.~\ref{algo1} with $\alpha\geq 1$ is 
\begin{align*}
& p_h^{r,i-1}(t)(g_h^{r,i}(t)-g_h^{r,i-1}(t))\geq \frac{c_h^r}{\alpha}(p_h^{r,i}(t)-p_h^{r,i-1}(t)),\\
& \forall i\in[I],t\in[T],h\in[H],r\in[R]\\
& q_k^{r,i-1}(t)(v_k^{r,i}(t)-v_k^{r,i-1}(t))\geq \frac{c_k^r}{\alpha}(q_k^{r,i}(t)-q_k^{r,i-1}(t)),\\
& \forall i\in[I],t\in[T],k\in[K],r\in[R]
\end{align*}
\end{definition}

The allocation-cost relationship shows that the cost in each time slot for scheduling a new job is bounded by the increase of the term $c_h^r p_h^r(t)$ and $c_k^r q_k^r(t)$ in (\ref{dualobj}). This is ensured by the update of the price function.

\begin{lemma}\label{thpdinequality}
If the allocation-cost relationship holds for $\alpha\geq 1$, then Alg.~\ref{algo1} ensures $P_i-P_{i-1}\geq \frac{1}{\alpha} (D_i - D_{i-1}),\forall i\in[I]$.
\end{lemma}
\begin{IEEEproof}
For any job $i\in[I]$, if job $i$ is rejected, then we have $P_i-P_{i-1}=D_i - D_{i-1}=0$ according to (\ref{outobj})(\ref{dualobj}), the inequality must hold.
If job $i$ is accepted with schedule $l$, {\em i.e.}, $x_{il}=1$. Then the increment value of the primal objective value $P_i$ is
\begin{align*}
P_i-P_{i-1}=f_i(t_{il}-a_i)
\end{align*}
Since $x_{il}=1$, according to Alg.\ref{algo1}, the constraint (\ref{outdualconstraint1}) is tight. Then we have
\begin{align*}
& f_i(t_{il}-a_i) = \mu_i +\sum_{t \in l} \sum_{h \in (t,l)} \sum_{r\in[R]} p_h^r(t) w_i^r y_{ih}^l(t)\\
& +\sum_{t \in l} \sum_{k \in (t,l)} \sum_{r\in[R]} q_k^r(t) s_i^r z_{ik}^l(t)\\
& = \mu_i + \sum_{t \in l} \sum_{h \in (t,l)} \sum_{r\in[R]} p_h^r(t) (g_h^{r,i}(t)-g_h^{r,i-1}(t))\\
& +\sum_{t \in l} \sum_{k \in (t,l)} \sum_{r\in[R]} q_k^r(t) (v_k^{r,i}(t)-v_k^{r,i-1}(t))
\end{align*}
Similarly, the increment value of the dual objective value $D_i$ is
\begin{align*}
& D_i-D_{i-1}= \mu_i +\sum_{t \in l} \sum_{h \in (t,l)}\sum_{r\in[R]}  (p_h^{r,i}(t) - p_h^{r,i-1}(t))c_h^r\\
& + \sum_{t \in l} \sum_{k \in (t,l)} \sum_{r\in[R]} (q_k^{r,i}(t) - q_k^{r,i-1}(t))c_k^r
\end{align*}
Summing up the allocation-cost relationship over all $t\in l$, $h \in (t,l)$, $k\in(t,l)$, $v \in [V]$, we have
\begin{align*}
P_i - P_{i-1} & \geq \frac{1}{\alpha} (D_i - D_{i-1} - \mu_i) + \mu_i\\
&= \frac{1}{\alpha} (D_i - D_{i-1}) + (1-\frac{1}{\alpha}) \mu_i
\end{align*}
As $\mu_i \geq 0$ and $\alpha \geq 1$, we have
\begin{align*}
P_i-P_{i-1}\geq \frac{1}{\alpha} (D_i - D_{i-1})
\end{align*}
\end{IEEEproof}

For specific $h\in[H]$, $r\in[R]$, we define $\alpha_h^r$ as the corresponding parameter in the allocation-cost relationship for any job $i\in[I]$ and any time slot $t\in[T]$. We also define $\alpha_k^r$ for specific $k\in[K]$, $r\in[R]$ in a similar way. Then $\alpha$ is just the maximum value of $\alpha_h^r$ and $\alpha_k^r$ among all $h\in[H], k\in[K], r\in[R]$. Without loss of generality, we assume that the resource demand of each worker or parameter server is much smaller compared to the capacity of that resource on one server, {\em i.e.}, $w_i^r \ll c_h^r$, $s_i^r \ll c_k^r$. This is common in real-world machine learning system as it is less likely that one worker/parameter server occupy a large percentage of resources in the whole server. As $g_h^r(t)$ ($v_k^r(t)$) increases from $0$ to $c_h^r$ ($c_k^r$), then we can claim that $\mathrm{d}g_h^r(t)=g_h^{r,i}(t)-g_h^{r,i-1}(t)$, $\mathrm{d}v_k^r(t)=v_k^{r,i}(t)-v_k^{r,i-1}(t)$, and derive a differential version of the allocation-cost relationship.

\begin{definition}\label{defdacr}
The differential allocation-cost relationship for Alg.~\ref{algo1} with $\alpha_h^r \geq 1$, $\alpha_k^r \geq 1$ is 
\begin{align*}
& p_h^r(t)\mathrm{d}g_h^r(t)\geq \frac{c_h^r}{\alpha_h^r}\mathrm{d}p_h^r(t), \forall t\in[T],h\in[H],r\in[R]\\
& q_k^r(t)\mathrm{d}v_k^r(t)\geq \frac{c_k^r}{\alpha_k^r}\mathrm{d}q_k^r(t), \forall t\in[T],k\in[K],r\in[R]
\end{align*}
\end{definition}

Next we show that a feasible $\alpha_h^r$ ($\alpha_k^r$) satisfies the differential allocation-cost relationship with price function $p_h^r(t)$ ($q_k^r(t)$) defined in (\ref{marginalcosts}).
\begin{lemma}\label{thalpha}
$\alpha_h^r=\ln \frac{U_1^r}{L_1}$, $\alpha_k^r=\ln \frac{U_2^r}{L_2}$ and the price functions defined in (\ref{marginalcosts}) satisfy the differential allocation-cost relationship.
\end{lemma}

\begin{IEEEproof}
The derivation of the marginal cost function is
\begin{align*}
\mathrm{d}p_h^r(t) & = p_h^{r'}(g_h^r(t)) \mathrm{d}g_h^r(t)\\
& = L_1\Big(\frac{U_1^r}{L_1}\Big)^{\frac{g_h^r(t)}{c_h^r}} \ln (\frac{U_1^r}{L_1})^{\frac{1}{c_h^r}}\mathrm{d}g_h^r(t)\\
\mathrm{d}q_k^r(t) & = q_k^{v'}(v_k^r(t)) \mathrm{d}v_k^r(t)\\
& = L_2\Big(\frac{U_2^r}{L_2}\Big)^{\frac{v_k^r(t)}{c_k^r}} \ln (\frac{U_2^r}{L_2})^{\frac{1}{c_k^r}}\mathrm{d}v_k^r(t)\\
\end{align*}
The differential allocation-cost relationship is
\begin{align*}
L_1\Big(\frac{U_1^r}{L_1}\Big)^{\frac{g_h^r(t)}{c_h^r}} \mathrm{d}g_h^r(t) \geq \frac{c_h^r}{\alpha_h^r} L\Big(\frac{U_1^r}{L_1}\Big)^{\frac{g_h^r(t)}{c_h^r}} \ln (\frac{U_1^r}{L_1})^{\frac{1}{c_h^r}}\mathrm{d}g_h^r(t)\\
L_2\Big(\frac{U_2^r}{L_2}\Big)^{\frac{v_k^r(t)}{c_k^r}} \mathrm{d}v_k^r(t) \geq \frac{c_k^r}{\alpha_k^r} L\Big(\frac{U_2^r}{L_2}\Big)^{\frac{v_k^r(t)}{c_k^r}} \ln (\frac{U_2^r}{L_2})^{\frac{1}{c_k^r}}\mathrm{d}v_k^r(t)
\end{align*}
which holds for $\alpha_h^r\geq \ln \frac{U_1^r}{L_1}$ and $\alpha_k^r\geq \ln \frac{U_2^r}{L_2}$.
Then we can set $\alpha=\max_{r\in[R]}(1, \ln \frac{U_1^r}{L_1}, \ln \frac{U_2^r}{L_2})$, which satisfies the differential allocation-cost relationship.
\end{IEEEproof}

\subsection{Proof of Theorem~\ref{thratio}}
\begin{IEEEproof}
According to Lemma~\ref{thalpha}, the marginal cost function used in Alg.\ref{algo1} satisfies the differential allocation-cost relationship with $\alpha=\max_{r\in[R]}(1, \ln \frac{U_1^r}{L_1}, \ln \frac{U_2^r}{L_2})$. Since the resource demand in a job $i$ is much smaller than the capacity, we can derive
\begin{align*}
& \mathrm{d}g_h^r(t) = g_h^{r,i}(t)-g_h^{r,i-1}(t)\\
& \mathrm{d}p_h^r(t) =  p_h^{r'}(g_h^r(t))(g_h^{r,i}(t)-g_h^{r,i-1}(t))=p_h^{r,i}(t)-p_h^{r,i-1}(t)\\
& \mathrm{d}v_k^r(t) = v_k^{r,i}(t)-v_k^{r,i-1}(t)\\
& \mathrm{d}q_k^r(t) =  q_k^{v'}(v_k^r(t))(v_k^{r,i}(t)-v_k^{r,i-1}(t))=q_k^{r,i}(t)-q_k^{r,i-1}(t)
\end{align*}

So the the differential allocation-cost relationship in Definition \ref{defdacr} implies the allocation-cost relationship in Definition \ref{defacr} holds for $\alpha=\max_{r\in[R]}(1, \ln \frac{U_1^r}{L_1}, \ln \frac{U_2^r}{L_2})$.

According to Alg.~\ref{algo1} and note that $\frac{1}{\eta_1} \leq \frac{\lceil E_i N_i M_i (\tau_i + 2e_i/b_i)\rceil \sum_{r\in[R]} w_i^r}{T\sum_{h\in[H]}\sum_{r\in[R]}c_h^r}$, then $\frac{T\sum_{h\in[H]}\sum_{r\in[R]}c_h^r}{\eta_1} \leq \lceil E_i N_i M_i (\tau_i + 2e_i/b_i)\rceil \sum_{r\in[R]} w_i^r,\forall i\in[I]$ is the minimum amount of overall resource consumption of workers of job $i$. Similarly, we have $\frac{1}{\eta_2} \leq \frac{\lceil E_i N_i M_i (\tau_i + 2e_i/b_i)\rceil \sum_{r\in[R]} s_i^r}{T\sum_{k\in[K]}\sum_{r\in[R]}c_k^r}$, and $\frac{T\sum_{k\in[K]}\sum_{r\in[R]}c_k^r}{\eta_2} \leq \lceil E_i N_i M_i (\tau_i + 2e_i/b_i)\rceil \sum_{r\in[R]} s_i^r,\forall i\in[I]$ is minimum amount of overall resource consumption of parameter servers of job $i$. We have
\begin{align}
& D_0 = \sum_{t\in[T]}\sum_{h\in[H]}\sum_{r\in[R]} L_1 c_h^r + \sum_{t\in[T]}\sum_{k\in[K]}\sum_{r\in[R]} L_2 c_k^r\nonumber\\
\end{align}
\begin{align}
& = \sum_{t\in[T]}\sum_{h\in[H]}\sum_{r\in[R]} \frac{1}{4\eta_1}\min_{i\in[I],l\in\mathcal{L}_i} \frac{f_i(t_{il}-a_i)}{\sum_{r\in[R]} \lceil E_i N_i M_i (\tau_i + 2e_i/b_i)\rceil w_i^r} c_h^r\nonumber\\
& + \sum_{t\in[T]}\sum_{k\in[K]}\sum_{r\in[R]} \frac{1}{4\eta_2}\min_{i\in[I],l\in\mathcal{L}_i} \frac{f_i(t_{il}-a_i)}{\sum_{r\in[R]} \lceil E_i N_i M_i (\tau_i + 2e_i/b_i)\rceil s_i^r} c_k^r\nonumber\\
& =  \frac{T\sum_{h\in[H]}\sum_{r\in[R]} c_h^r}{4\eta_1}\min_{i\in[I],l\in\mathcal{L}_i} \frac{f_i(t_{il}-a_i)}{\sum_{r\in[R]} \lceil E_i N_i M_i (\tau_i + 2e_i/b_i)\rceil w_i^r}\nonumber\\
& + \frac{T\sum_{k\in[K]}\sum_{r\in[R]} c_k^r}{4\eta_2}\min_{i\in[I],l\in\mathcal{L}_i} \frac{f_i(t_{il}-a_i)}{\sum_{r\in[R]} \lceil E_i N_i M_i (\tau_i + 2e_i/b_i)\rceil s_i^r}\nonumber\\
& \leq \frac{1}{4} \lceil E_i N_i M_i (\tau_i + 2e_i/b_i)\rceil \sum_{r\in[R]} w_i^r \min_{i\in[I],l\in\mathcal{L}_i}\frac{f_i(t_{il}-a_i)}{\sum_{r\in[R]} \lceil E_i N_i M_i (\tau_i + 2e_i/b_i)\rceil w_i^r}\nonumber\\
& +  \frac{1}{4} \lceil E_i N_i M_i (\tau_i + 2e_i/b_i)\rceil \sum_{r\in[R]} s_i^r \min_{i\in[I],l\in\mathcal{L}_i} \frac{f_i(t_{il}-a_i)}{\sum_{r\in[R]} \lceil E_i N_i M_i (\tau_i + 2e_i/b_i)\rceil s_i^r},\nonumber\\
& \forall i\in[I]\label{previneq}
\end{align}
We select $(i,l)=\arg\min_{i\in[I],l\in \mathcal{L}_i} f_i(t_{il}-a_i)$, then we have
\begin{align}
(\ref{previneq}) & \leq \frac{1}{4} \lceil E_i N_i M_i (\tau_i + 2e_i/b_i)\rceil \sum_{r\in[R]}  w_i^r \frac{f_i(t_{il}-a_i)}{\sum_{r\in[R]} \lceil E_i N_i M_i (\tau_i + 2e_i/b_i)\rceil w_i^r}\nonumber\\
& +\frac{1}{4} \lceil E_i N_i M_i (\tau_i + 2e_i/b_i)\rceil \sum_{r\in[R]} s_i^r \frac{f_i(t_{il}-a_i)}{\sum_{r\in[R]} \lceil E_i N_i M_i (\tau_i + 2e_i/b_i)\rceil s_i^r}\nonumber\\
& \leq \frac{1}{2} f_i(t_{il}-a_i)\nonumber\\
& \leq \frac{1}{2} OPT\label{aconejob}
\end{align}
where (\ref{aconejob}) is due to we assume the offline optimal solution accepts at least one job, which is reasonable in real-world machine learning system. Then we have $OPT\geq \min_{i\in[I],l\in\mathcal{L}_i} f_i(t_{il}-a_i)$.

According to Lemma~\ref{thpdratio} and Lemma~\ref{thpdinequality}, we conclude the proof.
\end{IEEEproof}
\end{appendices}}

\end{document}